\documentclass{pasj00}

\newcommand{\ha}{\mathrm{H}\alpha}
\newcommand{\flux}{\rm erg \ cm^{-2} s^{-1}}
\newcommand{\sfr}{M_{\odot}\mbox{ yr}^{-1}}

\begin{document}
\SetRunningHead{M. Sumiyoshi et al.}{Photometric H$\alpha$ and [O\emissiontype{II}] LF
of SDF and SXDF Galaxies: Implications for Future BAO Surveys} 
\Received{2000/12/31}
\Accepted{2001/01/01}

\title{Photometric H$\alpha$ and [O\emissiontype{II}] Luminosity
  Function of SDF and SXDF Galaxies: Implications for Future Baryon
  Oscillation Surveys}

\author{
Masanao \textsc{Sumiyoshi},\altaffilmark{1} 
Tomonori \textsc{Totani},\altaffilmark{1} 
Shunsuke \textsc{Oshige},\altaffilmark{1} 
Karl \textsc{Glazebrook},\altaffilmark{2} 
Masayuki \textsc{Akiyama},\altaffilmark{3} \\ 
Tomoki \textsc{Morokuma},\altaffilmark{4}
Kentaro \textsc{Motohara},\altaffilmark{5}
Kazuhiro \textsc{Shimasaku},\altaffilmark{6,7}
Masao \textsc{Hayashi},\altaffilmark{6} 
Makiko \textsc{Yoshida},\altaffilmark{6} \\
Nobunari \textsc{Kashikawa},\altaffilmark{4}
and Tadayuki \textsc{Kodama}\altaffilmark{4} 
}
\altaffiltext{1}{Department of Astronomy, Graduate School of Science,
Kyoto University, Kitashirakawa, Sakyo, Kyoto 606-8502} 
\email{sumiyosi@kusastro.kyoto-u.ac.jp}
\altaffiltext{2}{Centre for Astrophysics and Supercomputing, Swinburne
University of Technology, \\ Mail H39, PO Box 218, Hawthorn, VIC 3122,
Australia} 
\altaffiltext{3}{Astronomical Institute, Graduate School of Science, Tohoku 
University, Aramaki, Aoba, Sendai, Miyagi 980-8578}
\altaffiltext{4}{National Astronomical Observatory of Japan, 2-21-1
Osawa, Mitaka, Tokyo 151-8588}
\altaffiltext{5}{Institute of Astronomy, Graduate School of Science,
University of Tokyo, 2-21-1 Osawa, Mitaka, Tokyo 181-0015} 
\altaffiltext{6}{Department of Astronomy, Graduate School of Science,
University of Tokyo, Tokyo 113-0033}
\altaffiltext{7}{Reseach Center for the Early Universe, School of Science, 
University of Tokyo, 7-3-1 Hongo, Bunkyo, Tokyo 113-0033}



\maketitle


\begin{abstract}
  Efficient selection of emission line galaxies at $z \gtrsim 1$ by
  photometric information in wide field surveys is one of the keys for
  future spectroscopic surveys to constrain dark energy using the baryon
  acoustic oscillation (BAO) signature. Here we estimate the H$\alpha$ and
  [O\emissiontype{II}] line luminosity functions of galaxies at $z$ =
  0.5--1.7 using a novel approach where multi-wavelength imaging data is used
   to jointly estimate both photometric redshifts and star-formation rates.  
  These photometric estimates of line luminosities at high-redshift use the 
  large data sets of the
  Subaru Deep Field and Subaru XMM-Newton Deep Field (covering $\sim 1$
  deg$^2$) and are calibrated with the spectroscopic data of the local Sloan Digital Sky Survey
  galaxies.  The derived luminosity functions (especially H$\alpha$)
  are in reasonable agreement with the past estimates based on
  spectroscopic or narrow-band-filter surveys. This dataset is useful for
  examining the photometric selection of target galaxies for BAO
  surveys  because of the large cosmological volume covered and the large
  number of galaxies with detailed photometric
  information.  We use the sample to derive the photometric and physical properties of
  emission line galaxies to assist planning for future
  spectroscopic BAO surveys. We also show some examples of photometric
  selection procedures which can efficiently select these emission line
  galaxies.
\end{abstract}


\section{Introduction}

The precise measurements of the cosmological parameters in the past
decade have shown that we live in a Universe whose expansion is
accelerating \citep{Riess98,Perlmutter99,Spergel07,Komatsu08}.  
This has led to the paradigm of the  $\Lambda$CDM universe whose acceleration
is caused by a non-zero classical `cosmological
constant' ($\Lambda$), equivalent in effect to the energy density of
the vacuum in relativistic quantum mechanics. Although this simple model is almost perfectly consistent with the
cosmological data obtained so far, the value of $\Lambda$ is unexplained and in fact $10^{121}$ times
smaller than plausible physical values. These problems have motivate the generalisation of the 
accelerating component as a `dark energy'. The dark energy may be energy density
of a still unknown physical field, or modification of the theory of
gravity and spacetime on the cosmological scale may be required.  To
reveal the nature of dark energy has the potential to revolutionize
our understanding of the universe and fundamental physics. (See, e.g., 
\citet{Copeland06} for a theoretical review.) A key question is whether the dark
energy behaves exactly like $\Lambda$, or something more exotic. These
cases can be distinguished observationally by higher precision
measurements of the expansion history of the Universe.

One route to this expansion history is by using the baryon acoustic
oscillation (BAO) method which has attracted particular attention in
recent years as a promising new probe of dark energy
\citep{Blake03,Seo03,Glazebrook05}. The BAO scale in the Universe
serves as a `standard ruler' whose length is determined by the sound
horizon at the time of recombination and which imprints itself in the
clustering of matter as revealed by the cosmic microwave background
(CMB) at early times and the distribution of galaxies at later times.
Because the BAO scale ($\simeq 150$ comoving Mpc) is set by
straight-forward linear physics in the early Universe it is relatively
robust against astrophysical uncertainties and can be calibrated by
CMB observations \citep{EW04}. The BAO signatures have already been
seen in the distribution of low-redshift galaxies by the Two Degree
Field Galaxy Redshift Survey (2dFGRS) \citep{Cole05} and Sloan Digital
Sky Survey (SDSS) \citep{Eisenstein05,Tegmark06,Percival07,Okumura08}.

A clear next step is to detect BAO at higher redshifts, which would
yield constraints complementary to those of low redshift surveys.  A
number of such BAO surveys are planned, including both spectroscopic
and photometric surveys. In this paper we concentrate on spectroscopic
BAO surveys at $z \sim 1$, which are optimal surveys because apparent
BAO scales at $z \sim 1$ are the most sensitive to dark energy.
However BAO surveys need to cover minimum volumes of order 1
$h^{-3}$\,Gpc$^3$ in order to be effective \citep{Blake03} which
implies survey areas of at least 500--1000 deg$^2$.  Since the largest
fields of view on optical telescopes are of order 1 degree then for
surveys to be feasible exposure times must be kept short (below one
hour). Because of this star forming galaxies with emission lines are
of particular interest, since we can determine the redshifts with a
shorter exposure time. The emission lines useful for such surveys are
likely to be the [O\emissiontype{II}]$\lambda3727$ forbidden line
doublet in optical spectroscopy, and H$\alpha$ $\lambda6563$ in
near-infrared spectroscopy, both of which are well-known as
representative star formation indicators \citep{Kewley04,
  Moustakas06}.

Important issues for such BAO surveys are (1) whether there are a
sufficient number of emission line galaxies that can be detected with
a reasonable exposure time, and (2) how to select them from
photometric survey data as the targets of spectroscopic
observations. These issues can be examined by analysing the existing estimates
of the luminosity functions of emission line galaxies at $z \gtrsim
1$, based on slitless spectroscopic surveys (Yan et al. 1999; Hopkins
et al. 2000 in H$\alpha$), photometry-selected spectroscopic surveys
(Tresse et al. 2002 in H$\alpha$; Hogg et al. 1998; Teplitz et
al. 2003; Zhu et al. 2008 in [O\emissiontype{II}]), photometric
surveys using narrow-band filters (Geach et al. 2008 in H$\alpha$, Ly
et al. 2007; Takahashi et al. 2007 in [O\emissiontype{II}]), and
Fabry-Perot spectroscopic surveys (Hippelein et al. 2003 
in [O\emissiontype{II}]). 
However, the statistics of these studies (especially $\ha$) are rather limited,
because of the small survey area in spectroscopic surveys ($\lesssim$
60 arcmin$^2$ for $\ha$) and small redshift range in narrow-band
surveys ($\Delta z \sim$ 0.04).  In general BAO surveys are targeting
bright and low number-density galaxies compared with those targeted in
general studies of galaxy evolution, consequently limited statistics
may lead to significant uncertainties due to small statistics and
cosmic variances.  Recently \citet{Zhu08}, covering a large area
($\sim$14,000 galaxies in 2.45 deg$^{-2}$), has appeared but this is
the only large [O\emissiontype{II}] survey.

Here we study these issues based on wide-field broad-band photometric
data of the Subaru Deep Field (SDF) \citep{Kashikawa04} and the Subaru
XMM-Newton Deep Field (SXDF) \citep{Furusawa08}, with estimating
redshifts and emission line luminosities from star formation rate
based on the photometric redshift calculations. 
Although the redshift and emission line luminosity estimates are
less reliable than those of spectroscopic or narrow-band surveys, the
statistics is greatly improved by the SDF and SXDF data, which are the
deepest data among the surveys covering $\gtrsim$ 1
deg$^2$. Furthermore, deep photometric data in a variety of band
filters allow us to examine the photometric properties of emission
line galaxies in detail, to infer the required depth for imaging
surveys and efficient procedures to select targets for spectroscopic
observations.

We will present our results as far as possible in a general way so
that they are useful for any BAO survey planning. To complement this
we also examine the cases of specific BAO survey proposals.  Though
there are several BAO surveys planned at $z \sim 1$, we examine the
two proposed BAO surveys as examples.  One is the FastSound project,
which is a proposed BAO survey using the Fiber Multi-Object
Spectrograph (FMOS), which is a new multi-fiber NIR spectrograph for
the Subaru Telescope. FMOS has had engineering first light in May 2008
\citep{Iwamuro08} and is now undergoing further test observations.  A
BAO survey using H$\alpha$ emission line galaxies at $z \gtrsim 1$ has
been proposed \citep{Glazebrook04a}. The other
is a BAO survey based on [O\emissiontype{II}] emission line galaxies
using the Wide-field Fiber-fed Multi Object Spectrograph (WFMOS)
\citep{Bassett05}, which is a proposed next-generation optical
spectrograph for the Subaru Telescope having up to several thousand 
fibers.

This paper is organized as follows. In \S \ref{section: sample and photz} we
describe the photometric data sets used in this work and the
methodology of photometric redshift calculations.  In \S \ref{section: line calibration}, 
we estimate the emission line luminosities with
calibrations by the SDSS spectroscopic galaxy sample.  We then
calculate the H$\alpha$ and [O\emissiontype{II}] luminosity functions
and compare with those in previous studies, in \S \ref{section: line LF}.  In
\S \ref{section: properties_of_line_gal} we discuss implications for future
spectroscopic BAO surveys.  Finally we show some examples of color
selection of target galaxies for spectroscopic BAO galaxies in \S
\ref{section: color selection}. We summarize our results in \S
\ref{section: summary}. Throughout this paper, we use the standard cosmological
parameters of $H_{0}=70 \mbox{ km s}^{-1} \mbox{ Mpc}^{-1}$,
$\Omega_{m}=0.3$, $\Omega_{\Lambda}=0.7$, $\Omega_{b} =0.045$,
$\sigma_{8}=0.8$ in the $\Lambda$CDM universe \citep{Spergel07}.  All magnitudes are
given in the AB magnitude system.


\section{The Sample and Photometric Redshifts}\label{section: sample and photz} 

\subsection{The SDF and SXDF Galaxies}

\begin{longtable}{*{12}{c}}
\caption{Basic Information of SDF and SXDF} \label{field} 
\hline Field & Area [deg$^{2}$] & Number & 
\multicolumn{9}{c}{Limiting magnitudes\footnotemark[$*$] [mag]} 
\\  & & & $B$ & $V$ & $R_{\mathrm{C}}$ & $i'$ & $z'$ & $J$ & $K$ &
$3.6\micron$ & $4.5\micron$ \\
\hline
\hline
\endhead
SDF   &  0.114 &  17408  & 28.5 & 27.7 & 27.8 & 27.4 & 26.6 & 
24.0 & 24.3 & ----- & -----  \\
SXDF  & 0.732 &  76193  & 28.1 & 27.8 & 27.6 & 27.6 & 26.6 & 
24.5 & 24.0 &   23.1  &  22.4  \\
\hline
\multicolumn{8}{@{}l@{}}{\hbox to 0pt{\parbox{150mm}{\footnotesize
\par\noindent
\footnotemark[$*$] The limiting magnitudes are 3$\sigma$ at 2\arcsec
aperture for $B$, $V$, $R_{\mathrm{C}}$, $i'$, and $z'$ bands, and
3$\sigma$ at 3\arcsec.8 aperture for 3.6 and $4.5\micron$ bands. For
the $J$ and $K$ bands, they are 3$\sigma$ at 2\arcsec aperture (SDF) and
3$\sigma$ at 2\arcsec.3 aperture (SXDF).
} \hss}
}
\end{longtable}

The SDF is located at R.A. = 13$^{\mathrm{h}}$ 24$^{\mathrm{m}}$
38$^{\mathrm{s}}$.9, Decl. = +27\arcdeg~ 29\arcmin~ 25\arcsec.9 (J2000),
where the Galactic extinction is $E(B-V)=0.017$
\citep{Schlegel98}. All magnitudes presented below have been corrected
for the Galactic extinction. We use the $0.114 \ \mathrm{deg}^{2}$
region where both the Subaru/Suprime-Cam data in $BVRi'z'$ bands
\citep{Kashikawa04} and UKIRT/WFCAM data (\cite{Hayashi07};
\cite{Motohara09}, in preparation) in $JK$ bands are available.  The
limiting magnitudes of the optical and NIR data are shown in Table
\ref{field}. The seeing size of optical images is 0.98 arcsec FWHM
(1 pixel = 0.2 arcsec). We use 19,494 galaxies detected both 
in $i'$ and $K$ bands. The requirement for $K$-band photometry 
is because of the need for accurate photometric redshifts and
stellar contamination removal, however we will show below that this is not a significant 
source of bias.

The SXDF is centered at R.A. = 2$^{\mathrm{h}}$ 18$^{\mathrm{m}}$
00$^{\mathrm{s}}$.0 and Decl. = $-$5\arcdeg~ 00\arcmin~ 00\arcsec.0
(J2000), where the Galactic extinction is $E(B-V)=0.020$.  We use the
photometric data in $BVRi'z'$ bands \citep{Furusawa08} taken by
Subaru/Suprime-Cam, in $JK$ bands taken by UKIDSS survey
\citep{Warren07}, and in 3.6 and 4.5 $\mu$m bands taken by SWIRE
survey \citep{Lonsdale04}, in the area of $0.732 \ \mathrm{deg}^{2}$
where the data in all of these bands are available. The details of the
catalog will be reported elsewhere (\cite{Akiyama09}, in preparation).
The limiting magnitudes are given in Table \ref{field}, and the seeing
size of optical images is about 0.8 arcsec FWHM.  As in the SDF we
require $K$-band detection, and consequently we use 82,323 galaxies
detected both in $i'$ and $K$ bands.

Stellar objects are removed as follows.  For bright objects with $i' <
21.0$, we used the SExtractor \citep{Bertin96} 
stellarity and objects are removed when the $R$-band CLASS STAR
index is larger than 0.5.  For objects fainter than $i' = 21.0$, we
use the $Bz'K$ color-color diagram, and objects are removed if $(z'-K)
< 0.5 (B-z')-0.8$ and $(z'-K) < 0.3 (B-z')-0.4$.  (All galaxies
selected by the detection in $i'$- and $K$-bands are detected also in
$B$ and $z'$ bands.)  After the removal of stars by these procedures,
there remain 17,408 and 76,193 galaxies in the SDF and SXDF,
respectively.


\subsection{Photometric Redshift Estimate}

We use the publicly available code $hyperz$ \citep{Bolzonella00} for
our photometric redshift estimations.  We use 7 template models of
spectral energy distribution (SED) having exponentially decaying star
formation history with the characteristic time scales of $\tau = $
0.1, 0.3, 0.7, 1, 3, 5, and 15 Gyrs.  In addition to this, we added a
SED template with a constant star formation rate (SFR), and hence we
have 8 SED templates in total. We use the stellar spectrum library of
\citet{Bruzual03} with the solar metallicity ($Z=Z_{\odot}=0.02$) and
the initial mass function of \citet{Salpeter55} in the mass range of
0.1--100 $M_{\odot}$.  We apply the extinction law of
\citet{Calzetti00} for dust extinction ranging $A_{V}$ = 0.0--2.0 at
an interval of 0.2.  The stellar mass is estimated taking into account
instantaneous recycling with a typical value of the returned mass
fraction from evolved stars $R=0.25$ (e.g., Nagashima \& Yoshii 2004),
as $M_* = (1-R) M_{\rm SFR-int}$, where $M_{\rm SFR-int}$ is the
stellar mass calculated by the integration of SFR through the star
formation history of a galaxy.  It is well known that the Salpeter IMF
overestimates the number of low-mass stars compared to more realistic
IMFs, We use the Salpeter IMF as it is convenient to use with
$hyperz$, however we divide our final masses by 1.82 to convert them
to the \authorcite{BG03} IMF \citep{BG03} which is more realistic but has a particularly
simple scaling with respect to the Salpeter IMF \citep{Glazebrook04b}.

\begin{figure}
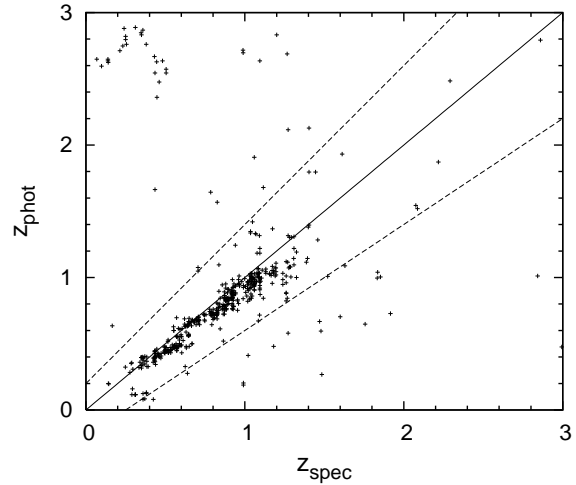

\begin{center}
\FigureFile(80mm,80mm){fig1_comp_z.eps}
\end{center}
\caption{
Comparison of spectroscopic and photometric redshifts of SXDF
galaxies. The solid and dashed lines show the lines of $z_{\rm phot} = z_{\rm spec}$ 
and $|z_{\rm phot} - z_{\rm spec}|/(1+z_{\rm spec}) = 0.2$, respectively. 
}
\label{z_spec-z_phot}
\end{figure}

\begin{figure}
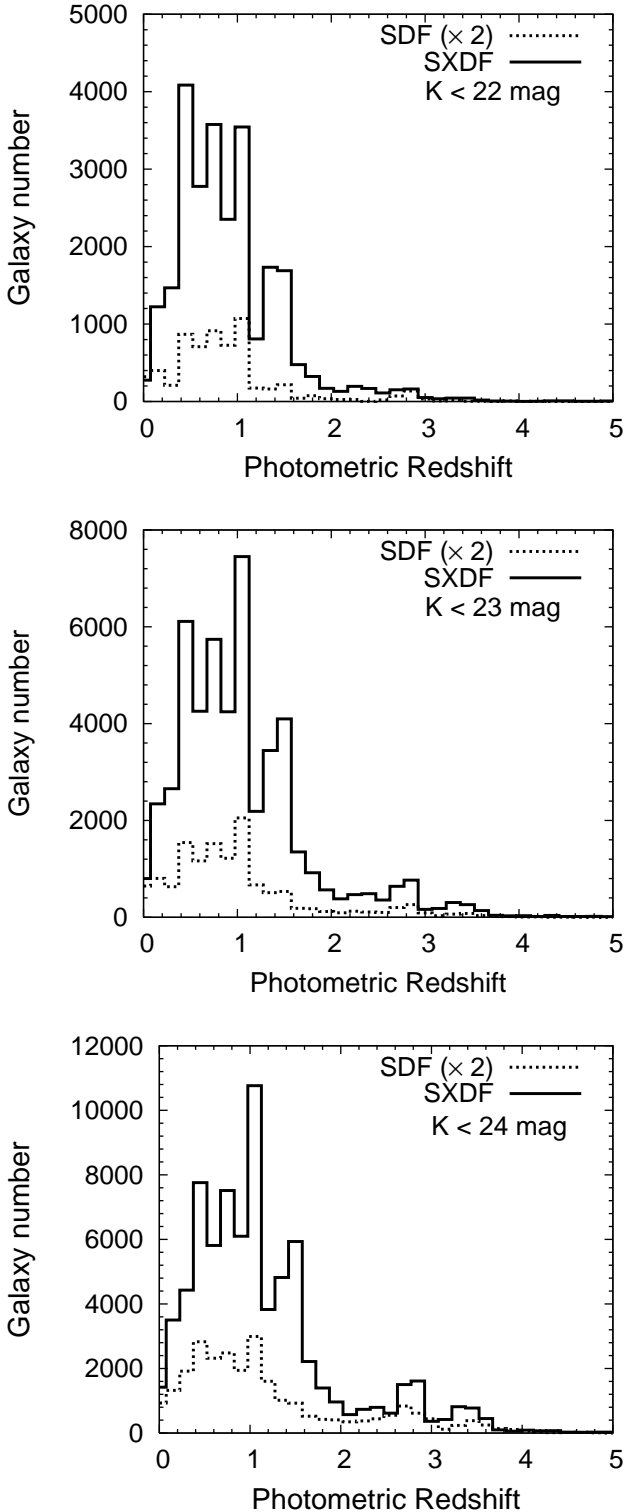

\begin{center}
\FigureFile(88mm,80mm){fig2a_z_histogram_K22.eps}
\FigureFile(88mm,80mm){fig2b_z_histogram_K23.eps}
\FigureFile(88mm,80mm){fig2c_z_histogram_K24.eps}
\end{center}
\caption{
The photometric redshift distributions of 
SDF and SXDF galaxies brighter than the $K$-magnitude 
of 22 (top), 23 (middle), and 24 mags (bottom). 
The number of SDF galaxies multiplied by 2 is shown in each panel. 
}
\label{z_histogram}
\end{figure}

The photometric redshifts ($z_{\mathrm{phot}}$) calculated in this way
are compared with the spectroscopic redshifts ($z_{\mathrm{spec}}$) of
550 X-ray undetected galaxies in SXDF having observed spectra (Figure
\ref{z_spec-z_phot}).  As in \citet{Ilbert06}, we define the redshift
accuracy as the median of $1.48|\Delta z|/(1+z)$, and outlier fraction
as the fraction of those having $|\Delta z|/(1+z) > 0.2$.
The redshift accuracy and outlier fraction in $z_{\rm
  phot} < 2.0$ and $z_{\rm spec} < 2.0$ are 0.030 and 6\%. This level
of agreement is sufficient for the purposes of this work for most of
the galaxies.  It can be seen that galaxies at $z_{\rm spec} \sim 0.3$
tend to be estimated as $z_{\rm phot} \sim 3.0$, because of confusion
between the 4000$\rm \AA$ and Lyman breaks. However, since we are
considering objects in the range $0.5 < z < 1.7$, this confusion will
not affect our results.  The photometric redshift distributions of SDF
and SXDF galaxies in the different thresholds of $K$-magnitude are
given in Figure \ref{z_histogram}.

In the following sections, we will mainly show the result of SXDF galaxies,   
because the photometric redshift calculations in SXDF are
expected to be more reliable than those in SDF due to the advantage of having 
mid-infrared data.

\section{Line Luminosity Calibration by SDSS Galaxies}\label{section: line calibration} 

We estimate emission line luminosities from SFRs of the best-fit
SED templates in the photometric redshift calculations.  Our primary goal is not to
measure the SFRs themselves but rather use them as an intermediate step towards 
the final line luminosities.
Accordingly rather than use standard conversion
factors from line luminosity to SFR (e.g., \cite{Kennicutt98}), we calculate 
our own conversion factors as part of
our calibration process. This ensures we empirically obtain the best estimate 
of the relation between
line luminosities and our photometrically estimated SFR, 
including systematic errors such as dust and metallicity effects. The
calibration is done by using the large
spectroscopic sample of local galaxies of the SDSS.


\subsection{The SDSS Sample}

We use the $u'g'r'i'z'$ photometric catalog publicly released from the
SDSS studies at Max-Planck-Institute for Astrophysics / Johns Hopkins
University\footnote{http://www.mpa-garching.mpg.de/SDSS/}, which is
based on the fourth SDSS data release \citep{Adelman-McCarthy06}.  We
select a sample with $14.5 < r_{\mathrm{petro}} < 17.77$ and $0.023 <
z < 0.399$ in order to detect $\ha$ and [O\emissiontype{II}] lines in
the wavelength coverage of SDSS, where $r_{\mathrm{petro}}$ is the
Petrosian magnitude of $r$-band. We use the sample of star-forming
galaxies constructed by \citet{Brinchmann04}, and the SDSS data set
used in this work includes 75,757 objects, whose mean redshift is 0.081.

The SDSS photometric zero points are close to the AB convention but
they are not reproduced exactly.  We convert the SDSS photometry
system into the AB system using the AB correction of
\citet{Eisenstein06}.  The SDSS spectroscopic observations cannot
measure the total line emissions from galaxies, because of the limited
size of the fiber diameter 3\arcsec.  We correct this aperture
effect by the relation between Petrosian and fiber magnitudes as
described in \citet{Hopkins03}.


\subsection{Calibration}

We use the $hyperz$ on the SDSS $u'g'r'i'z'$ catalog for our photometric 
redshift estimations, as in the case of SDF and SXDF. 
In this work, we are interested in estimating observable line
luminosities, rather than the
true extinction-corrected line luminosity corresponding to SFR. 
Therefore we compare the
SFR that is estimated by $hyperz$ but reduced by the magnitudes
corresponding to extinction ($A_{\rm H\alpha}$ and $A_{\rm
  [O\emissiontype{II}]}$), with the observed line luminosities
($L_{\rm H\alpha, obs}$ and $L_{\rm [O\emissiontype{II}], obs}$).  The
amount of extinction is calculated from the $A_V$ estimates by
$hyperz$ and the adopted extinction law.  

\begin{figure}
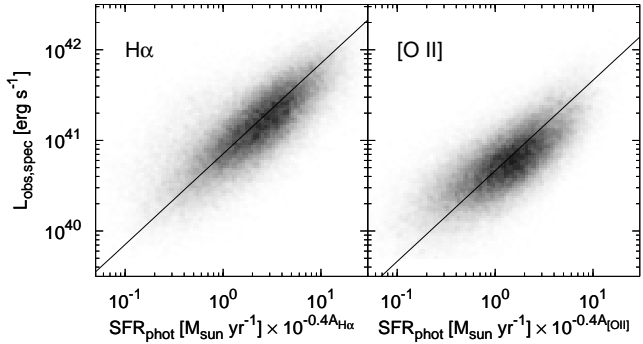

\begin{center}
\FigureFile(88mm,88mm){fig3_SDSS_Line.eps}
\end{center}
\caption{ The extincted SFRs estimated by photometric redshift
  calculations versus H$\alpha$ and [O\emissiontype{II}] line
  luminosities measured by spectroscopic data of SDSS galaxies. The
  solid lines are the best-fit linear relation.  
  The scatters in the case of H$\alpha$ and [O\emissiontype{II}] 
  are $\pm$0.24dex and $\pm$0.35dex, respectively. }
\label{line_calib} 
\end{figure}

The correlation between the two quantities of the SDSS galaxies is
shown in Fig. \ref{line_calib}, for H$\alpha$ and [O\emissiontype{II}]
lines.  It should be noted that we have corrected SFR estimates by an
additional factor of $[d_L(z_{\rm spec})/d_L(z_{\rm phot})]^2$, to
allow for the difference between the true spectroscopic redshift and
that returned by $hyperz$.  For low-redshift samples such as SDSS a
small redshift difference $\Delta z$ does not significantly change the
observed SED shape (as $\lambda \Delta z$ is much smaller than the
band width) and hence the $hyperz$ fit. However it has a much stronger
effect on derived luminosities which is removed by this correction.
We cross-checked this method of SFR estimates using a sub-sample of a
few hundred SDSS galaxies where we manually adjusted $hyperz$ to
return the SED best fit at $z_{\rm spec}$. We found the SFRs from the
two approaches consistent to within a factor of two. We use the former
approach as it is much more efficient for large samples with the
existing $hyperz$ codebase.

We see a tight correlation between SFR and H$\alpha$ luminosities, 
and we obtain the best-fit conversion relation as: 
\begin{eqnarray*} 
\log \left( \frac{L_{\ha, \rm obs}}{[\mbox{erg s}^{-1}]}\right)
= \log \left( \frac{\rm SFR_{phot}}{[\sfr ]} \right) - 0.4A_{\ha} \\+ 40.85. 
\end{eqnarray*} 
The offset $+40.85$ is different from that of \citet{Kennicutt98} by
0.25 dex, which is not unreasonable considering the uncertainties in
stellar spectrum libraries and efficiency of ionization photon
production.

\begin{figure}
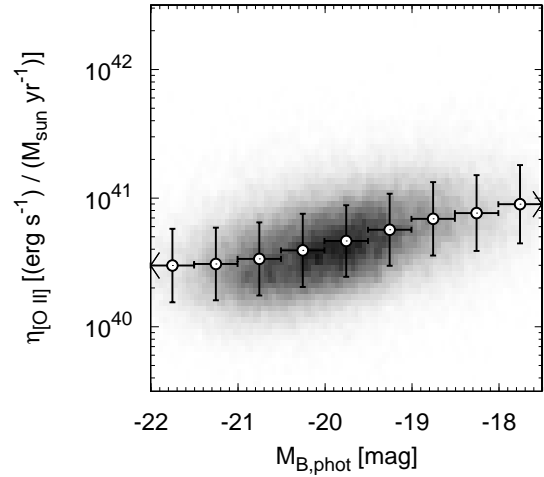

\begin{center}
\FigureFile(88mm,88mm){fig4_OII_factor.eps}
\end{center}
\caption{ The conversion factor $\eta _{\rm [O\emissiontype{II}]}$
  from SFR to [O\emissiontype{II}] line luminosity as a function of
  absolute $B$-band magnitude, $M_{B, \rm phot}$. The small dots are
  the SDSS galaxies. The open circles and errors show the mean
  and standard deviation of the SDSS galaxy distribution in each
  $M_{B, \rm phot}$ bin. 
}
\label{OII_factor} 
\end{figure}

\begin{figure}
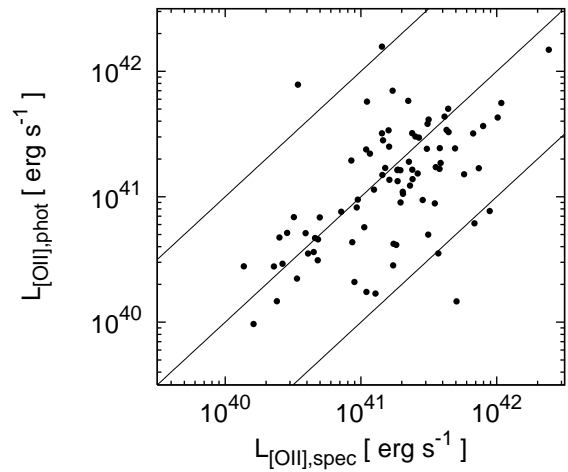

\begin{center}
\FigureFile(80mm,80mm){fig5_comp_FOII.eps}
\end{center}
\caption{
Comparison of spectroscopic and photometric [O\emissiontype{II}] luminosities of SXDF
galaxies with available spectra. The solid lines show the lines of
$\log{(L_{\rm [O\emissiontype{II}], phot} / L_{\rm [O\emissiontype{II}], spec})}
= 0$ and $\pm 1.0$. The scatter is approximately $\pm$0.5dex.
}
\label{comp_FOII}
\end{figure}

Next we consider the [O\emissiontype{II}] emission. As can be seen in
the right panel of Fig. \ref{line_calib}, the correlation with SFR is
not as good as that of H$\alpha$ emission, which is almost certainly due to the effect of
metallicity and extinction \citep{Kewley04, Moustakas06}. Hence, following Moustakas et al.,  we
calibrate the conversion law from SFR to [O\emissiontype{II}]
luminosities as a function of absolute $B$-band magnitude ($M_{B, \rm
  phot}$) in order to remove the dominant luminosity dependence of these effects. Figure \ref{OII_factor} shows the
conversion factor, $\eta _{\rm [O\emissiontype{II}]}$, versus absolute
$B$-band magnitude, $M_{B, \rm phot}$.  Here the conversion factor is
defined by the following equation:
\begin{eqnarray*} 
  \log \left( \frac{L_{\rm [O\emissiontype{II}], obs}}{[\mbox{erg s} ^{-1}]} 
 \right) 
  = \log \left( \frac{\rm SFR_{phot}}{[\sfr]} \right) 
  - 0.4A_{\rm [O\emissiontype{II}]} 
  \\+ \log[ \eta _{\rm [O\emissiontype{II}]}(M_{B, \rm phot}) ]. 
\end{eqnarray*} 
We derived
$\eta_{\rm [O\emissiontype{II}]}$ from calculating best fit values
using SDSS data in every $M_{B, \rm phot}$ bins, which are shown by
open circles in Fig. \ref{OII_factor}.  The best-fit formula of the
conversion factor is given as follows:
\begin{eqnarray*} 
\log [ \eta _{\rm [O\emissiontype{II}]}(M_{B, \rm phot}) ] = 43.24 + 0.13 M_{B, \rm phot}.
\end{eqnarray*} 

One possible criticism of this calibration is that it is based on local
galaxies and hence might not applicable to high-$z$ galaxies (for
example due to evolution in metallicity effects). We test this issue
by using 88 SXDF galaxies with [O\emissiontype{II}] luminosities
estimated from the emission line equivalent widths, and Figure
\ref{comp_FOII} shows the comparison between photometric and
spectroscopic [O\emissiontype{II}] luminosity estimates.  Here we
again correct the photometric line luminosity estimates by the factor
of $[d_L(z_{\rm spec})/d_L(z_{\rm phot})]^2$ as for the SDSS sample,
to remove the error coming from the distance estimate. The agreement
between the two is reasonable though the scatter is significant
($\sim$0.5dex).  We also note also that Moustakas et al. showed that
their approach for [O\emissiontype{II}] to SFR conversion worked
reasonably well for galaxies at $z\simeq 1$.


\section{Line Luminosity Functions}\label{section: line LF}


\subsection{H$\alpha $ Luminosity Function}

\begin{figure}
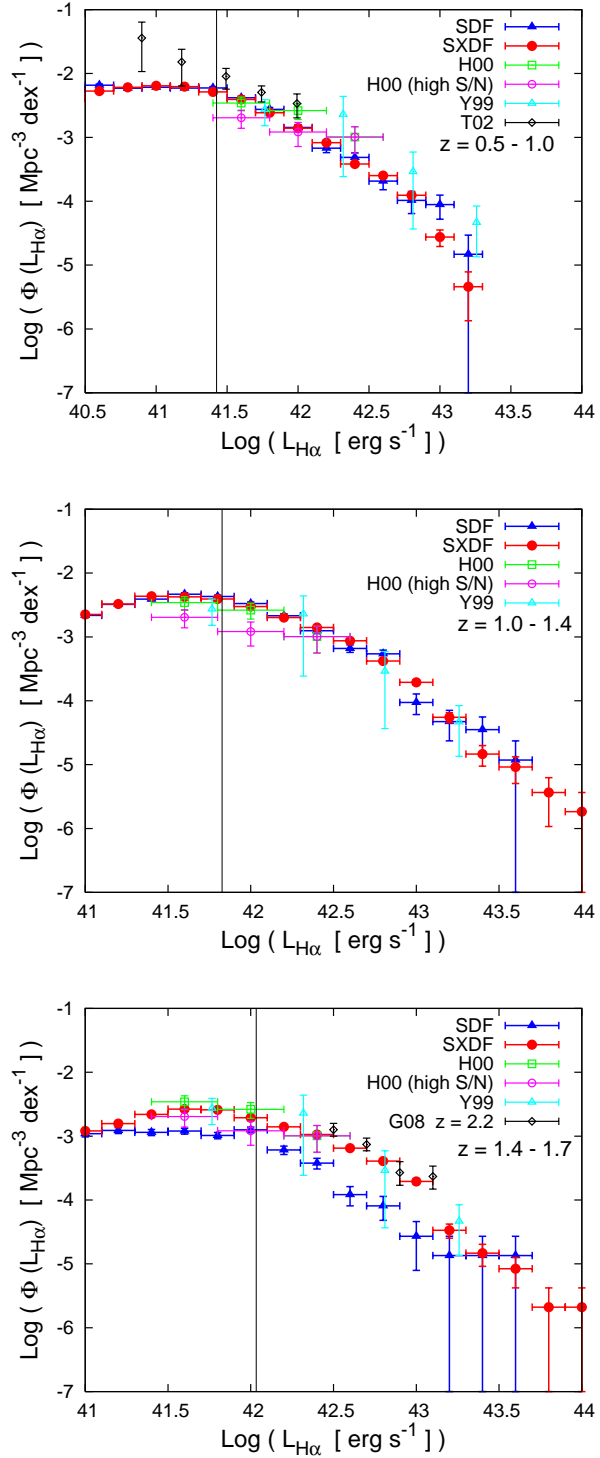

  \begin{center}
    \FigureFile(80mm,80mm){fig6a_LF_Ha_z=0.5-1.0.eps}
    \FigureFile(80mm,80mm){fig6b_LF_Ha_z=1.0-1.4.eps}
    \FigureFile(80mm,80mm){fig6c_LF_Ha_z=1.4-1.7.eps}
  \end{center}
  \caption{ The $\ha$ luminosity functions of SDF galaxies (filled
    triangles) and SXDF galaxies (filled circles) in $z$ = 0.5--1.0
    (top), $z$ = 1.0--1.4 (middle), and $z$ = 1.4--1.7 (bottom).  The
    vertical lines indicate the luminosities corresponding to $F_{\ha}
    = 10^{-16} \flux$ at each redshift interval.  Estimates by
    previous studies are also plotted, whose symbols are indicated in
    the figure.  The redshift ranges of the previous estimates are as
    follows: \authorcite{Hopkins00} (\yearcite{Hopkins00}, H00) in $z$
    = 0.7--1.8, \authorcite{Yan99} (\yearcite{Yan99}, Y99) in $z$ =
    0.7--1.9, \authorcite{Tresse02} (\yearcite{Tresse02}, T02) in $z$
    = 0.5--1.1, and \authorcite{Geach08} (\yearcite{Geach08}, G08) in
    $z \sim$ 2.23, even though there is a small mismatch of redshift
    range.  The \lq\lq high S/N \rq\rq data of \citet{Hopkins00} are
    using only galaxies with spectroscopic confirmation or high S/N
    ratios. The data of G08 are obtained by a narrowband
    observation, while the others are spectroscopic observations. }
\label{Ha_LF}
\end{figure}

\begin{figure}
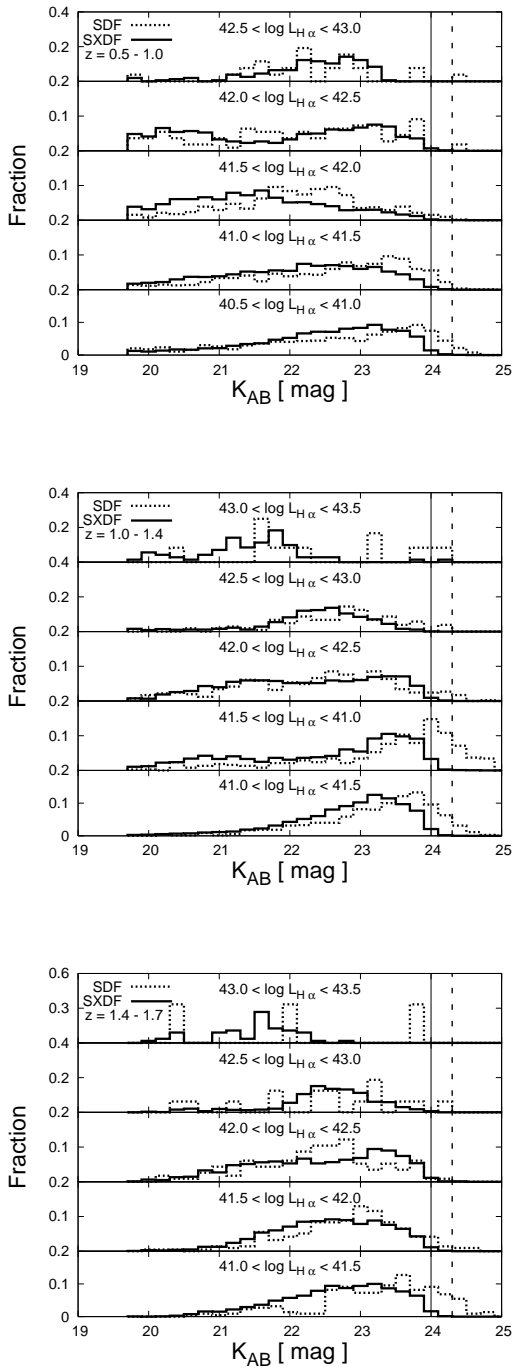

  \begin{center}
    \FigureFile(88mm,88mm){fig7a_Ha_K_mag_z=0.5-1.0.eps}
    \FigureFile(88mm,88mm){fig7b_Ha_K_mag_z=1.0-1.4.eps}
    \FigureFile(88mm,88mm){fig7c_Ha_K_mag_z=1.4-1.7.eps}
  \end{center}
  \caption{ The $K$-magnitude distributions of SDF (dotted) and SXDF
    galaxies (solid) in five $L_{\ha}$ (in erg/s) bins at $z$ =
    0.5--1.0 (top), 1.0--1.4 (middle), and 1.4--1.7 (bottom). The
    vertical lines are the $K$ limiting magnitudes of SDF (dashed
    line, 3$\sigma$ at 2'' aperture) and SXDF (solid line, 3$\sigma$
    at 2''.3 aperture). 
   }
\label{Ha_K_mag}
\end{figure}

We construct the $\ha$ luminosity function per unit logarithmic
luminosity interval in a given redshift interval $(z_1, z_2)$, by
simply dividing the observed number of galaxies by the comoving volume
corresponding to the redshift interval. We do not use the more
sophisticated $1/V_{\max}$ method (e.g., \cite{Yan99}) to estimate
LFs, because we cannot determine a clear limiting flux of H$\alpha$
emission for galaxies originally selected by broad band fluxes.
Instead, we discuss the effect of limiting magnitudes on the LF
estimates below.

The derived H$\alpha$ luminosity functions from SDF data (filled
triangles) and SXDF data (filled circles) are shown in Figure
\ref{Ha_LF}, for three different redshift ranges of $0.5 < z < 1.0$
(top), $1.0 < z < 1.4$ (middle), and $1.4 < z < 1.7$ (bottom).  We
also plot the $\ha$ luminosity functions measured by previous studies,
after correcting the cosmological parameter into those used in this
work.  The redshift ranges of the previous studies are slightly
different from ours (see figure caption), and we plot the previous
data in the panel of similar redshift ranges in this figure.

To examine the effect of limiting magnitudes on the LF results, the
$K$-magnitude distributions of galaxies in several bins of $L_{\ha}$
are plotted in Figure \ref{Ha_K_mag}. Since the $K$-band observation
is relatively shallower than the $i'$ band, we can examine the effect
of the limiting magnitudes by comparing these distributions with the
$K$-magnitude limits of SDF and SXDF. We find that the effect of
limiting magnitude is not significant at $L_{\rm H\alpha} \gtrsim
10^{41.4}, 10^{41.8}$, and $10^{42.0}$ erg/s 
for the redshift ranges of
$z=$ 0.5--1.0, 1.0--1.4, and 1.4--1.7, respectively, which are
corresponding to the H$\alpha$ flux of $F_{\ha} \geq 10^{-16} \flux$. 
These luminosities are indicated by the vertical
solid lines in Figure \ref{Ha_LF}.  Because galaxies for the BAO
surveys considered in this work are brighter than this H$\alpha$ flux,
the incompleteness does not seriously affect the main results of this
work.  As another test, we estimate the number of $\ha$ emission
galaxies that are brighter than $F_{\ha} = 10^{-16} \flux$ 
but have been missed in our sample because of the requirement
of $K$-band detection, by using the $i'$ selected sample.  In this
case, we cannot use the $Bz'K$ diagram to remove stars and hence
stellar contamination may increase, but we find that the number of
bright emission galaxies increases at most 20\% from our baseline
analysis requiring $K$-band detection.

The agreement with the previous data derived by spectroscopic or
narrow-band filter observations is reasonably good above the limiting
magnitudes, providing some credence to our study based on photometric
line luminosity estimates.  However, the brighter-end of the LF at $z
=$ 1.4--1.7 estimated by the SDF data is significantly lower than
those by the SXDF data and previous studies, which cannot be explained
solely by the limiting magnitude effect.  We find that the reason for
this discrepancy is confusion between galaxies at $z \sim 0$ and $z \gtrsim 1.4$
due to the lack of the mid-infrared data in 3.6 and 4.5
$\mu $m bands in the SDF. We confirm this by checking that the LFs estimated by SXDF
galaxies without using mid-infrared data are similar to those by SDF
galaxies, thus the SXDF LF is preferred in the bottom panel of Figure \ref{Ha_LF}.


\subsection{[O\emissiontype{II}] Luminosity function}

\begin{figure}
  \begin{center}
    \FigureFile(80mm,80mm){fig8a_LF_OII_z=0.5-1.0.eps}
    \FigureFile(80mm,80mm){fig8b_LF_OII_z=1.0-1.4.eps}
    \FigureFile(80mm,80mm){fig8c_LF_OII_z=1.4-1.7.eps}
  \end{center}
  \caption{ The same as Fig. \ref{Ha_LF}, but for [O\emissiontype{II}]
    line luminosity functions.  The vertical lines indicate the
    luminosities corresponding to $F_{[\rm O\emissiontype{II}]} \sim
    10^{-16.5} \flux$ on each redshift interval. 
    The redshift ranges of
    the data of previous studies are as follows:
    \authorcite{Hippelein03} (\yearcite{Hippelein03}, H03) at $z \sim
    0.88$ (top panel) and $\sim 1.18$ (middle); \authorcite{Ly07}
    (\yearcite{Ly07}, L07) at $z \sim 0.89$ (top), $z \sim 1.19$
    (middle), and $z \sim 1.47 $ (bottom); \authorcite{Takahashi07}
    (\yearcite{Takahashi07}, T07) at $z \sim 1.19 $ (middle); 
    \authorcite{Zhu08} (\yearcite{Zhu08}, Z08) at $z$ = 0.752--0.926 (top), 
    $z$ = 1.108--1.277 (middle), and $z$ = 1.277--1.446 (bottom). 
    The data of H03 and Z08 are obtained by spectroscopic observations, 
    while those of L07 and T07 are based on narrowband observations. }
\label{OII_LF} 
\end{figure}

\begin{figure}
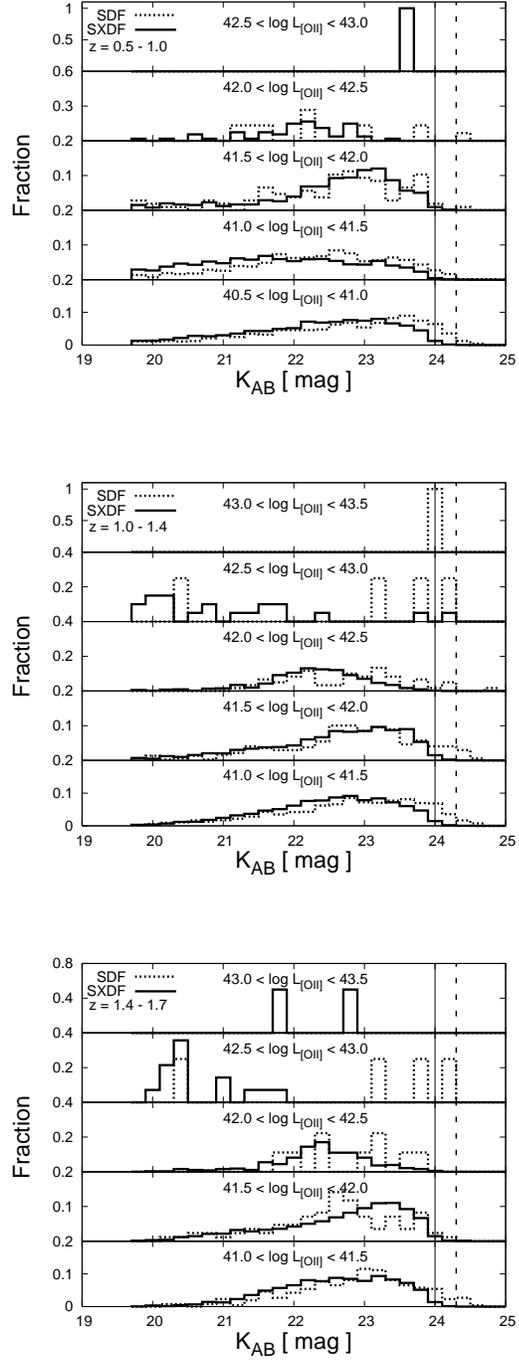

  \begin{center}
    \FigureFile(88mm,88mm){fig9a_OII_K_mag_z=0.5-1.0.eps}
    \FigureFile(88mm,88mm){fig9b_OII_K_mag_z=1.0-1.4.eps}
    \FigureFile(88mm,88mm){fig9c_OII_K_mag_z=1.4-1.7.eps}
  \end{center}
  \caption{ The same as Fig. \ref{Ha_K_mag}, but for
    [O\emissiontype{II}] line luminosity functions.  
}
\label{OII_K_mag} 
\end{figure}

We also calculate the [O\emissiontype{II}]$\lambda$3727 luminosity
function in SDF and SXDF, which are shown in Fig. \ref{OII_LF}, as
well as those by previous studies (see figure caption).  The limiting
magnitude effect is analyzed in the same way as with the H$\alpha$
luminosity function, by breaking it down in $K$-magnitude bins (Fig. \ref{OII_K_mag}). 
The figure shows that the limiting magnitude effect is not significant for $L_{[\rm
  O\emissiontype{II}]} \gtrsim 10^{41.9}$, $10^{41.3}$, and $10^{41.5}$
erg/s at $z = $ 0.5--1.0, 1.0--1.4, and 1.4--1.7, respectively, which
are corresponding to $F_{[\rm O\emissiontype{II}]} \geq 10^{-16.5} \flux$. 
These luminosities are also shown by the
vertical lines in Figure \ref{OII_LF}.  As in the analysis of the
H$\alpha$ LF, we estimate the number of [O\emissiontype{II}] emission
galaxies brighter than $F_{[\rm O\emissiontype{II}]} = 10^{-16.5} \flux$ 
using the sample selected by $i$-band detection
regardless of the $K$-band detection, and we find that the increase of
emission galaxies is at most $\sim$ 30\%.

Although the results from different studies are reasonably consistent
at $z$ = 0.5--1.0 and 1.0--1.4, the agreement is rather worse than for
the H$\alpha$ LF.  The reason could be due to the large scatter of the
$L_{\rm [O\emissiontype{II}]}$--SFR relation or uncertainties in our
calibration procedure.  The paucity of galaxies at the faint end of
our LF at $z$ = 1.0--1.4 compared with previous studies is most likely
due to the limiting magnitude effect. However, an even larger
discrepancy is found at the faint-end of the LF at $z =$ 1.4--1.7,
which can not be explained solely by the limiting magnitude effect.
(The disagreement between the [O\emissiontype{II}] LFs of SDF and SXDF
is again caused by the lack of the mid-infrared data in SDF.)  One
possibility is the contamination by different redshift emission line
objects in the sample of \citet{Ly07} which was selected by
narrow-band photometry. However according to Ly et al. the
contamination rate of their [O\emissiontype{II}] sample, estimated
from their spectroscopic sample, is only 10--20\%. The cosmic variance
due to the small survey volume of Ly et al. may also have caused the
discrepancy.  Further systematic uncertainties in our photometric line
luminosity estimates, such as the photometric redshift errors at $z >
1.4$ (see Fig. \ref{z_spec-z_phot}), may also be the origin of the
discrepancy.  Since the Ly et al. data is the only previous study that
can be compared with the faint-end of our LF in this redshift range
more data from independent studies are highly desirable.


\section{Properties of Emission Line Galaxies}\label{section: properties_of_line_gal} 

In this section we discuss implications for future BAO surveys, based
on the emission line luminosity estimates obtained above. We intend to
make our results as general as possible, but a specific discussion
assuming existing proposals for future BAO surveys is also useful and 
illustrative. In particular, we discuss the proposed BAO surveys 
using FMOS and WFMOS 
as examples of $\ha$ and [O\emissiontype{II}] surveys, respectively. 
We consider three representative redshift ranges in the following analysis: 
$0.5 < z < 1.0$, $1.0 < z < 1.4$, and $1.4 < z < 1.7$, as used in the LF
estimates.  The whole redshift range $0.5 < z < 1.7$ roughly
corresponds to the range that can be probed by optical
[O\emissiontype{II}] and near-infrared H$\alpha$ surveys.


\subsection{Galaxy Surface Density as Functions of Line Flux}

\begin{figure}
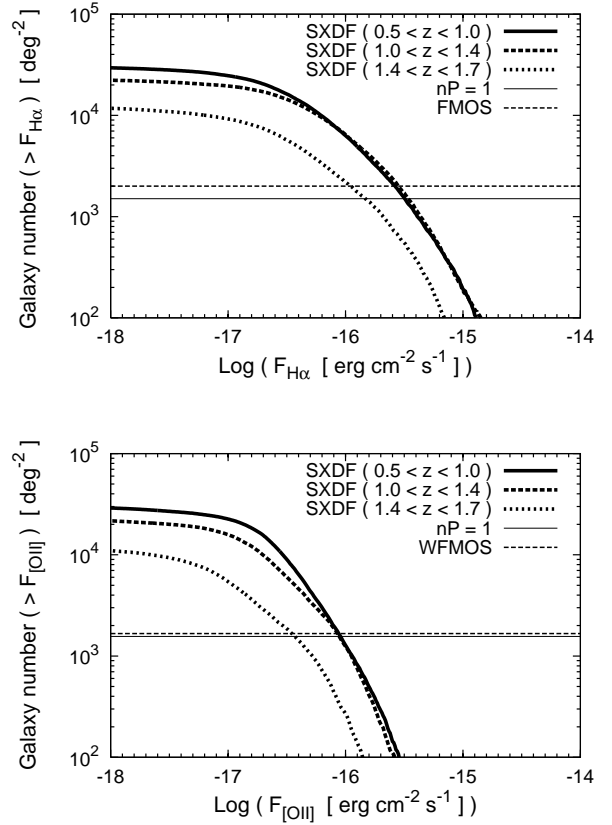

  \begin{center}
    \FigureFile(80mm,80mm){fig10a_SXDS_flux_density_Ha.eps}
    \FigureFile(80mm,80mm){fig10b_SXDS_flux_density_OII.eps}
  \end{center}
  \caption{ The cumulative number counts of galaxies as functions of
    H$\alpha$ (top) and [O\emissiontype{II}] (bottom) flux. The solid,
    dashed, and dotted lines are for the redshift ranges of $0.5 < z <
    1.0$, $1.0 < z <1.4$, and $1.4 < z <1.7$, respectively. The solid
    horizontal lines indicate the surface number density corresponding
    to $nP = 1$ at $z \sim 1$. (This density does not change
    significantly for the three redshift ranges.)  The dashed
    horizontal lines represent the FMOS and WFMOS fiber density in the
    top and bottom panels, respectively.  }
\label{flux_density}
\end{figure}

\begin{figure*}
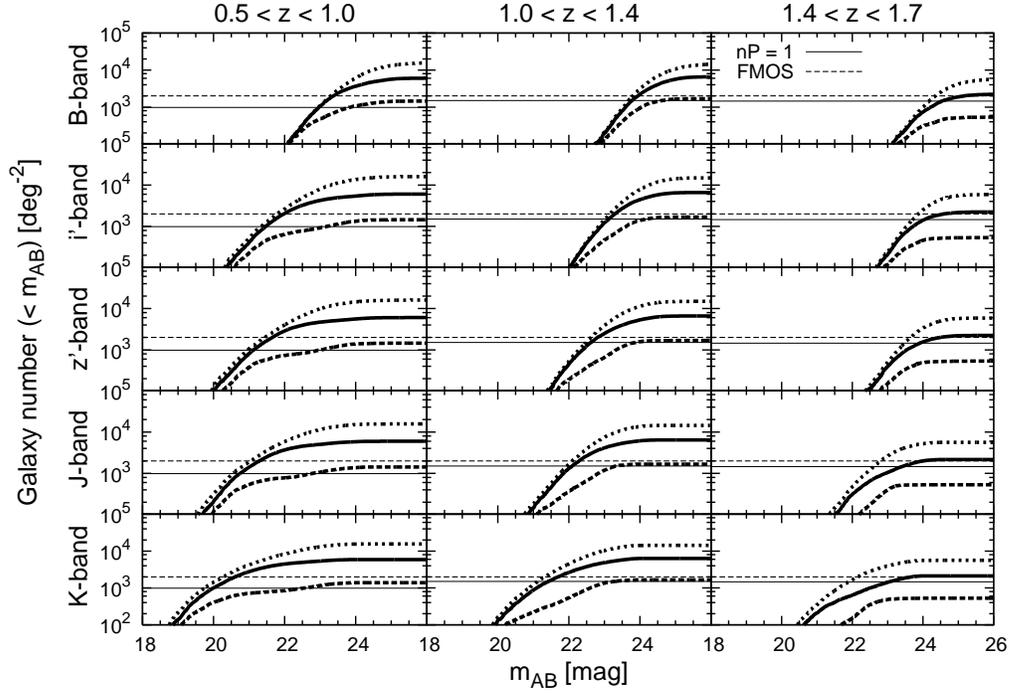

\begin{center}
\FigureFile(135mm,135mm){fig11_SXDS_Ha_mag_dens.eps}
\end{center}
\caption{ The cumulative galaxy numbers of the $\ha$ emitting galaxies
  in SXDF as a function of magnitudes in popular band filters. They
  are shown separately for the three redshift intervals as
  indicated. The dashed, solid, and dotted lines show the number of
  emission line galaxies brighter than the three different threshold
  $\ha$ fluxes of $10^{-15.5}$, $10^{-16}$, and $10^{-16.5} \flux$,
  respectively. The horizontal lines (solid and dashed) indicate the
  surface number density corresponding to $nP = 1$ at each redshift interval 
  and the FMOS fiber density, respectively.  }
\label{Ha_magfunc}
\end{figure*}

\begin{figure*}
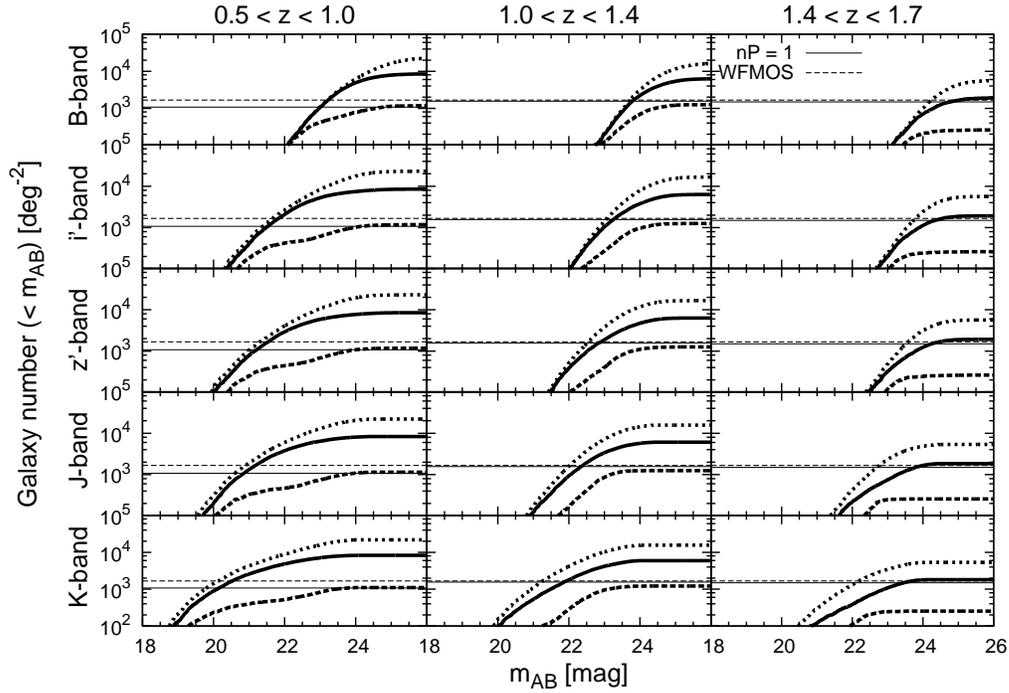

\begin{center}
\FigureFile(135mm,135mm){fig12_SXDS_OII_mag_dens.eps}
\end{center}
\caption{ The same as the Figure \ref{Ha_magfunc}, but for the
  [O\emissiontype{II}] line. The dashed, solid and dotted lines are
  for the three different threshold [O\emissiontype{II}] fluxes of
  $10^{-16}$, $10^{-16.5}$, and $10^{-17} \flux$, respectively.  The
  horizontal lines (solid and dashed) indicate the surface number
  density corresponding to $nP = 1$ at each redshift interval and the WFMOS fiber
  density, respectively.  }
\label{OII_magfunc} 
\end{figure*}

First we check the available number of emission line galaxies in the
three redshift intervals.  These are the maximum numbers of emission
line galaxies that can be used in any BAO survey.  We plotted the
cumulative number of SXDF galaxies as a function of $\ha$ and
[O\emissiontype{II}] flux in Figure \ref{flux_density}. The solid
horizontal line in the panel indicates the surface number density of
$nP = 1$ assuming $z \sim 1$, where $n$ is galaxy number density and
$P$ is the amplitude of the power spectrum of galaxy density
fluctuation at $k = 0.20 h \rm Mpc^{-1}$, which is roughly the
linear-nonlinear transition scale at $z \sim 1$.  The survey power is
maximized by choosing $nP = 1$ when the observing time is fixed
\citep{Seo03,Glazebrook05}. Here we used the formula of
\citet{Eisenstein98} to calculate the CDM power spectrum. The linear
clustering biases $b = 1.19$ and 1.17 are taken into account ($P_{\rm
  gal} = b^2 P_{\rm CDM}$) for $\ha$ and [O\emissiontype{II}] lines,
respectively. These values are from the bias estimates for emission
line galaxies at $1.0 < z < 1.4$ by stellar mass estimates, as
described in \S \ref{section: bias}.

We define $F_{\ha,\mathrm{min}}$ as the flux above which the number of
galaxies is equal to the surface number density corresponding to $nP
=1$, and hence $F_{\ha,\mathrm{min}}$ gives the minimally required
flux sensitivity to perform any BAO survey by emission line galaxies.
We find that these values are $\log{(F_{\ha,\mathrm{min}}/[\flux])}=$
$-$15.5, $-$15.5 and $-$15.8 for the redshift intervals of $0.5 < z <
1.0$, $1.0 < z < 1.4$ and $1.4 < z< 1.7$, respectively. These fluxes
are brighter than the FMOS sensitivity, $\sim \log{(F_{\ha}/[\flux])}
= -16.0$ for 1 hour integration \citep{Eto04}, and hence there are a
enough number of H$\alpha$-emitting galaxies for a BAO survey by FMOS.
The dotted horizontal line represents the fiber density of FMOS (400
in 0.2 deg$^{2}$), which is close to the galaxy number density
corresponding to $nP = 1$, meaning that FMOS is an efficient
instrument for a BAO survey than can collect a necessary number of
galaxy spectra by one time observation per field, although BAO was not
the original scientific target of this instrument.

Similarly, we define $F_{\rm [O\emissiontype{II}], min}$ for
[O\emissiontype{II}] line, and we find that these are $\log{(F_{\rm
    [O\emissiontype{II}], min}/[\flux])}= -16.1$, $-16.1$ and $-16.4$
for the three redshift intervals, respectively, which should be
compared with the line sensitivity of WFMOS, $\log{(F_{\rm
    [O\emissiontype{II}]}/[\flux])} \sim -16.5$, for 0.5 hour
integration \citep{Parkinson07}.  The horizontal dotted line is the
planned fiber density of WFMOS (3000 in 1.8 deg$^{2}$). As in the case
of H$\alpha$ line for FMOS, there is a sufficient number of emission
line galaxies that can be used for a [O\emissiontype{II}] BAO survey
by WFMOS.  

Finally we compare this result with that of the WiggleZ survey, a BAO
survey targeting emission-line galaxies in the redshift range of $0.2
< z < 1.0$ \citep{Glazebrook07,Blake09}. 
The number of galaxies brighter than $F_{\rm [O\emissiontype{II}]}$ =
10$^{-16} \flux$ at $z=$0.6--0.7 by our estimates is 210 deg$^{-2}$,
while that of actually observed by the WiggleZ survey is 35
deg$^{-2}$.  The paucity of galaxy number of the WiggleZ is likely due
to strong UV and $r$-band selection criteria of the survey.

\subsection{Galaxy Surface Density as Functions of Broad-band Filter 
Magnitudes}

Even if there are a sufficiently large number of galaxies for BAO
surveys, they must be pre-selected by imaging surveys, and the
required depth for such photometric selection is crucial in planning
BAO surveys. Figure \ref{Ha_magfunc} shows the cumulative galaxy
number of the $\ha$ emitting galaxies in SXDF as a function of
magnitudes in popular band filters. They are shown separately for the
three redshift intervals and different H$\alpha$ threshold fluxes as
indicated.  Figure \ref{OII_magfunc} is the same as Figure
\ref{Ha_magfunc}, but for [O\emissiontype{II}] and WFMOS.

From these figures, one can infer the necessary (but not sufficient)
depth in each band filter for photometric selection of targets.  For
example, we need imaging survey limits deeper than $B = 23.9$, $i' =
23.3$, $z' = 22.8$, $J = 22.3$, and $K = 21.7$ to detect BAO target
galaxies brighter than the H$\alpha$ sensitivity of FMOS ($>
10^{-16.0} \ \rm erg \ cm^{-2} s^{-1}$) with a number density of FMOS
fibers at $1.0 < z < 1.4$.  Similarly, the required imaging depths for
the WFMOS [O\emissiontype{II}] sensitivity ($>10^{-16.5} \ \rm erg \
cm^{-2} s^{-1}$) become $B = 23.8$, $i' = 23.3$, $z' = 22.9$, $J =
22.4$, and $K = 21.9$.  The similar magnitudes for FMOS and WFMOS
indicate that the BAO targets for these two surveys have similar SFRs.
While $\ha$ line luminosity is generally larger than
[O\emissiontype{II}], NIR observations are less sensitive than optical
observations, and hence the required integration time for Subaru
becomes comparable for H$\alpha$ observed by FMOS and
[O\emissiontype{II}] observed by WFMOS for a given SFR.


\subsection{Stellar Mass, Extinction, and Size}

\begin{figure}
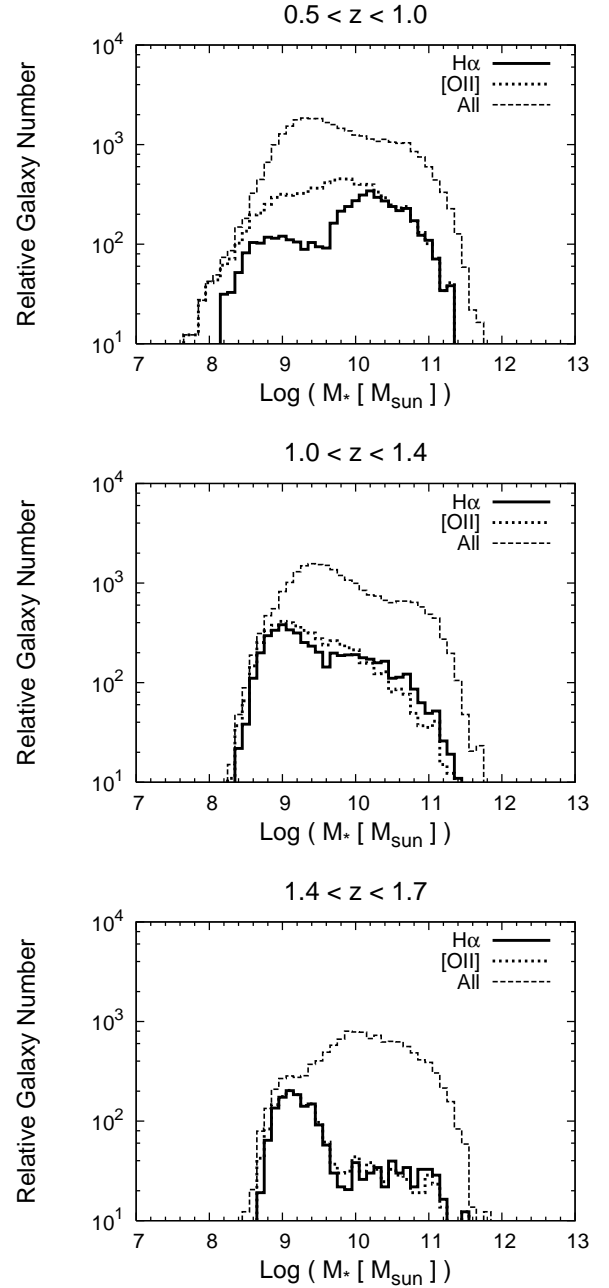

\begin{center}
\FigureFile(80mm,80mm){fig13a_SXDS_Mass_z_0510.eps}
\FigureFile(80mm,80mm){fig13b_SXDS_Mass_z_1014.eps}
\FigureFile(80mm,80mm){fig13c_SXDS_Mass_z_1417.eps}
\end{center}
\caption{ The stellar mass distributions of $\ha$ and
  [O\emissiontype{II}] emission galaxies in SXDF at $z=$0.5--1.0
  (top), 1.0--1.4 (middle), and 1.4--1.7 (bottom). The $\ha$ and
  [O\emissiontype{II}] threshold line fluxes are $10^{-16}$ and
  $10^{-16.5}$ $\flux$, respectively. We also show the mass
  distributions of all galaxies in the SXDF sample in the same
  redshift range, for the comparison with the results of emission line
  galaxies.  }
\label{Mass_flux} 
\end{figure}

Here we discuss some physical properties of the emission line galaxies
that are useful for a planning of a BAO survey.  First, we show the
stellar mass distribution of SXDF galaxies that have been calculated
in the photometric redshift estimates, in Figure \ref{Mass_flux} with
the threshold fluxes of $10^{-16}$ $\flux$, for H$\alpha$, or
$10^{-16.5}$ $\flux$, for [O\emissiontype{II}] flux, in the three
redshift intervals. We compare the mass distributions of these results
with those of all galaxies in SXDF, and then we find that almost all
galaxies which have small masses are emission line galaxies.  The
logarithmic means of the stellar mass distributions are presented in
Table \ref{mass}, for several different line flux thresholds.  The
mean stellar mass does not change significantly with the threshold
line flux in a fixed redshift interval, and does not change either
with redshift for a fixed threshold flux. The mean stellar mass of
line emitting galaxies is rather low, but we note that there is quite a
large spread of stellar masses from very massive galaxies down to
dwarfs.  This result indicates that the line luminosity is not
strongly correlated with stellar mass of galaxies.

\begin{longtable}{*{4}{c}}
\caption{Logarithmic means of the stellar mass of SXDF galaxies}
\label{mass} \hline Line flux [ $\flux$ ]& 
\multicolumn{3}{c}{$\log M_*$ [ M$_{\odot}$ ]\footnotemark[$*$] } \\  &
$0.5 < z < 1.0$ & $1.0 < z < 1.4$ & $1.4 < z < 1.7$ \\ 
\hline
\hline
\endhead
$\log F_{\ha} > -16.5$   &  9.8 	&  9.7 	& 9.9 	\\ 
$\log F_{\ha} > -16.0$   &  9.9 	&  9.6 	& 9.6 	\\ 
$\log F_{\ha} > -15.5$   &  9.7 	&  9.4	& 9.6 	\\ 
\hline
$\log F_{[\rm O \emissiontype{II}]} > -17.0$   &  9.6 	&  9.6 	& 9.8 	\\ 
$\log F_{[\rm O \emissiontype{II}]} > -16.5$   &  9.7 	&  9.5 	& 9.5 	\\ 
$\log F_{[\rm O \emissiontype{II}]} > -16.0$   &  9.5 	&  9.3 	& 9.7 	\\ 
\hline
\multicolumn{4}{@{}l@{}}{\hbox to 0pt{\parbox{100mm}{
\footnotesize 
\par\noindent 
\footnotemark[$*$] See Fig. \ref{Mass_flux} for the presentation of
stellar mass distributions for threshold fluxes of $\log F_{\rm
  H\alpha} > -16.0$ and $\log F_{\rm [O\emissiontype{II}]} > -16.5$.
}\hss}}
\end{longtable}

\begin{figure}
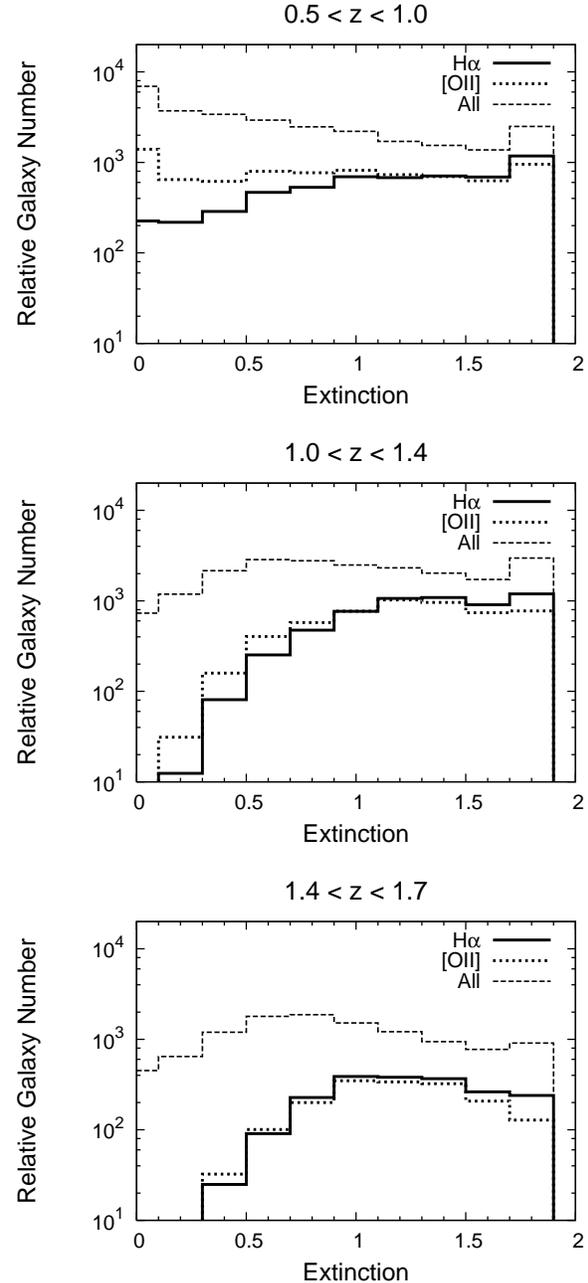

\begin{center}
\FigureFile(80mm,80mm){fig14a_SXDS_Av_z_0510.eps}
\FigureFile(80mm,80mm){fig14b_SXDS_Av_z_1014.eps}
\FigureFile(80mm,80mm){fig14c_SXDS_Av_z_1417.eps}
\end{center}
\caption{ The same as Fig. \ref{Mass_flux}, but for
the extinction ($A_V$) distributions of $\ha$ and
  [O\emissiontype{II}] emission galaxies.}
\label{Av_flux} 
\end{figure}

Next we examine the extinction ($A_V$) distribution of SXDF galaxies
estimated by the photometric redshift method in the three redshift
ranges, as shown in Figure \ref{Av_flux}.  The extinction is
distributed with a large scatter, with mean values of $A_V = 1.2$ and
emission line galaxies have larger extinctions in this figure.
However the distribution of galaxies selected by [O\emissiontype{II}]
at the lowest redshift range of $0.5 < z < 1.0$ shows a significant
number of galaxies with $A_V \sim 0$ compared with other redshift
ranges.  It is uncertain whether this is real or caused by systematic
uncertainties of photo-$z$ calculation.

\begin{figure}
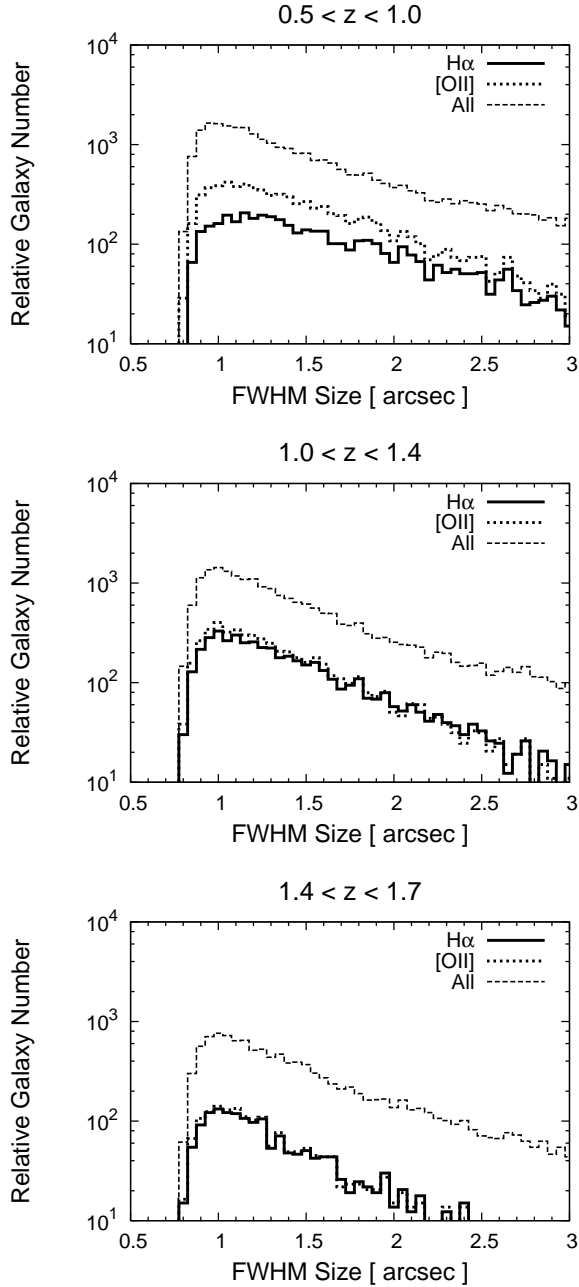

\begin{center}
\FigureFile(80mm,80mm){fig15a_SXDS_FWHM_z_0510.eps}
\FigureFile(80mm,80mm){fig15b_SXDS_FWHM_z_1014.eps}
\FigureFile(80mm,80mm){fig15c_SXDS_FWHM_z_1417.eps}
\end{center}
\caption{ The same as Fig. \ref{Mass_flux}, but for the $B$-band FHWM
  size distributions of $\ha$ and [O\emissiontype{II}] emission
  galaxies.  The seeing is not deconvolved in these sizes.  }
\label{FWHM_flux} 
\end{figure}

We also calculate the histogram of FWHM sizes of the SXDF galaxies in
the three redshift ranges, as shown in Figure \ref{FWHM_flux}.  One
can compare these sizes with the fiber aperture or slit width in a
planned survey to estimate the loss of emission line light due to the
extended size of galaxies. The FWHM size was measured in the $B$ band,
because the galaxy profile in the shortest wavelength is expected to
be close to that in emission lines.  Seeing ($\sim 0.8$ arcsec FWHM)
has not been subtracted, meaning the values correspond to the actual
image size under realistic observing conditions.  In Figure
\ref{FWHM_flux}, every distribution of FWHM size has a lower cut-off
at approximately 0.75 arcsec and gradually decays with increasing
size. The lower cut-off is due to the original image quality of the
optical images in SXDF.  For reference, the fiber diameter of FMOS is
1\arcsec.2 diameter, and the planned fiber diameter of WFMOS is also
about 1\arcsec. These results indicate that a significant fraction of
emission line flux could be lost out of the fibers in large size
galaxies, and this effect must be taken into account in BAO survey
planning.

\subsection{Linear Bias}
\label{section: bias}

\begin{figure}
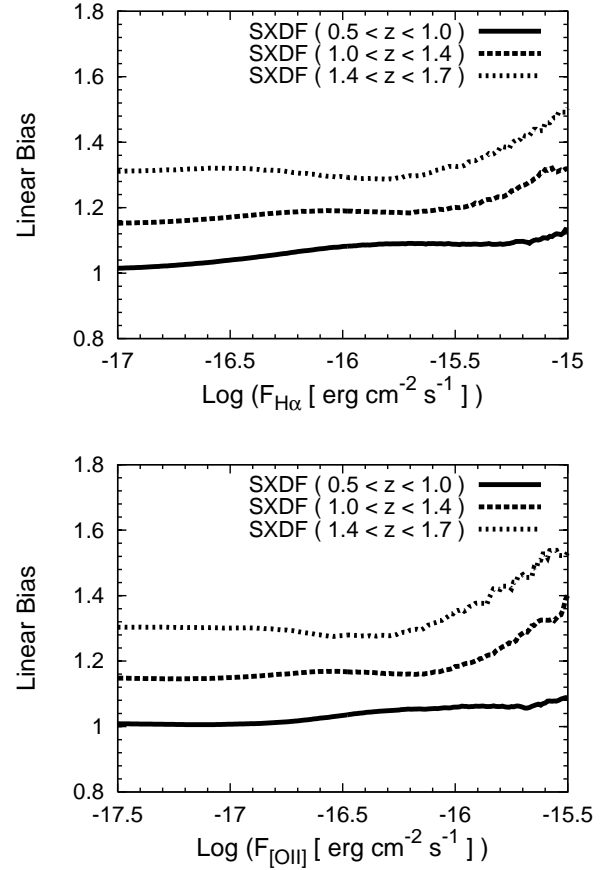

\begin{center}
\FigureFile(80mm,80mm){fig16a_SXDS_bias_FHa.eps}
\end{center}
\begin{center}
\FigureFile(80mm,80mm){fig16b_SXDS_bias_FOII.eps}
\end{center}
\caption{
The top panel shows the linear bias estimated for the SXDF galaxies that
are brighter than a given H$\alpha$ flux, for the three redshift
intervals as indicated in the panel. The bottom panel is the same as the
top, but for the [O\emissiontype{II}] line.
} 
\label{bias_flux} 
\end{figure}

The bias of clustering is important to estimate the detectability of
BAO signatures. We therefore calculate linear bias of clustering by
using the baryon mass estimate of galaxies in SXDF.  The baryon mass
is estimated from the photometric redshift calculations, including
stellar mass and gas mass that is not yet converted into stars. 
(In the case of the SED template of constant SFR, 
we suppose that gas mass is equal to the stellar mass estimated 
from the phot$-z$ calculations.) 
The estimated mean gas mass is consistent with that of 
UV-selected galaxies at $z \sim 2$ obtained by \citet{Erb06}.  
We then make a rough estimate of dark matter halo mass, assuming the
cosmic baryon-to-DM ratio, $\Omega_b/\Omega_{\rm DM}$:
\begin{eqnarray*} 
M_{\mathrm{DM}} = \frac{\Omega_{\rm DM}}{\Omega_{b}} (M_* + M_{\rm gas}), 
\end{eqnarray*} 
where $\Omega_{\rm DM} = \Omega_{m} - \Omega_{b}$, and $\Omega_{m}$
and $\Omega_{b}$ are the density parameters for matter and baryons,
respectively.  We then calculate the logarithmic mean of dark halo
mass for a given threshold line flux in a given redshift interval, and
then calculate linear bias from the estimated mass using the bias
model of dark matter haloes of \citet{Cooray02}. 
The biases estimated in this way
are shown in Figure \ref{bias_flux} for the SXDF galaxies as functions
of the threshold H$\alpha$ and [O\emissiontype{II}] fluxes.  The
estimated linear bias is typically $b \sim $1--1.5.  Since the stellar
mass of galaxies does not change with the threshold flux or redshift,
the linear bias does not depend strongly on the threshold flux, but it
becomes larger at higher redshifts.

We estimated the linear bias assuming that the DM-to-baryon mass ratio
of galaxies is the same as the cosmic value, but it may not be the
case, and it is likely an underestimation because of the following
reasons. First, the gas mass estimated by the star formation history
of photo-z templates (exponential SFR evolution) is highly uncertain,
and gas mass can be much larger than stellar mass in young,
line-emitting starburst galaxies (Erb et al. 2006). Second, we have
implicitly assumed that all the baryons are either in stars or in cold
gas participating in star-formation. However, a significant amount of
diffuse hot gas could also lie in dark halos. It has been found
empirically that the ratio of dark halo and stellar mass is
approximately 2--20$\Omega_{\rm DM} / \Omega_{b} $ for different
luminosities of galaxies at $z \sim 2$ (e.g., \cite{Hayashi07}).
Therefore we also calculate the biases with a 3 and 10 times larger DM
mass than the above estimates, but the linear biases are only modestly
increased by 10\% and 30\%, respectively.

We can compare the bias estimates with those by clustering analysis in
the literature at $z \simeq $1.0--2.0, by converting correlation
length into bias of line-emission galaxies.  
\citet{Blake09} obtained $b \sim 1.6$ for [O\emissiontype{II}]
emitters at $z = $0.5--1.0 from the real space correlation function of
the WiggleZ project data \citep{Glazebrook07,Blake09}, with a
threshold [O\emissiontype{II}] flux of 1 hour integration on the 3.9m
Anglo-Australian Telescope ($\sim 10^{-16} \flux$).  This bias is
larger than our estimate, and this can not be explained by increasing
the DM-to-baryon mass ratio of galaxies.  However, this is likely to
be due to the strong additional $r$-band selection criteria of the
WiggleZ survey ($20 < r < 22.5$).  We found that, when we add $20 < r
< 22.5$, our bias estimate is increased by 20\% as well.  Therefore,
our bias estimate is in rough agreement with the WiggleZ clustering
data when the increase of our bias prediction by larger $\Omega_{\rm
  DM} / \Omega_{b}$ values is taken into account.  \citet{Geach08}
obtained $b=2.0$ for $\ha$ emitters at $z \sim $2.2 by angular
correlation function, with a threshold H$\alpha$ flux of $\sim
10^{-16}$ erg s$^{-1}$ cm$^{-2}$. This is out of the redshift range
that we investigated, but it seems also roughly consistent with our
estimate considering the increasing trend of bias with redshift.


\subsection{Clustering Property}

\begin{figure}
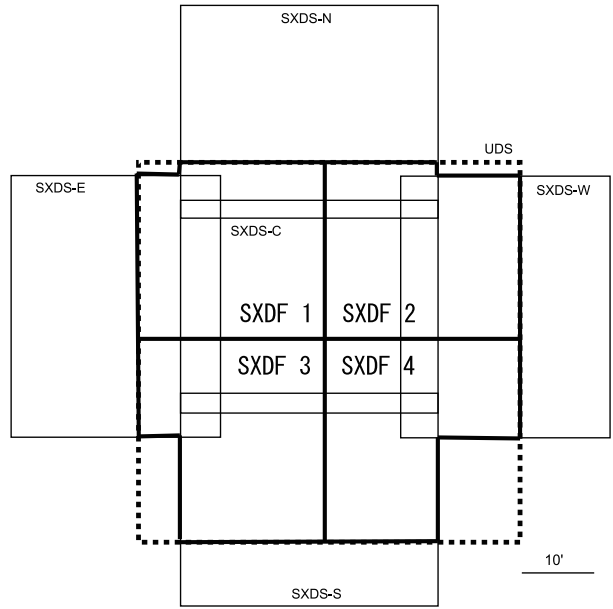

\begin{center}
\FigureFile(80mm,80mm){fig17_sxdf_sub.eps}
\end{center}
\caption{ Overview of SXDF.  The regions indicated by thin solid lines
  are the five fields of view of Subaru/Suprime-Cam (SXDS-C, N, S, W,
  and E).  The region shown by dashed lines is the area where NIR data
  by UKIRT/WFCAM is available.  All of the field is covered by
  mid-infrared observations by Spitzer/IRAC.  We used the overlapping
  region of the optical and NIR data, and the thick solid lines show
  the four sub-fields defined in this work.  }
\label{sxdf_sub} 
\end{figure}

Next we examine the field-to-field variation of the observed number of
emission-line galaxies in SXDF, and compare it with those expected
from the linear bias estimated above and the CDM structure formation
theory. A detailed study on the clustering properties by using angular
correlation function will be reported elsewhere. We divided the region
in SXDF used in this work (the region having both optical and NIR
data) into four sub-regions (SXDF 1--4) with roughly the same area of
$\sim 0.18 \ \rm deg^2$ (see Figure \ref{sxdf_sub} and Table
\ref{field-to-field}).

\begin{longtable}{*{9}{c}}
\caption{Field-to-field variation of galaxy number}
\label{field-to-field} \hline Field & Area [deg$^2$] & 
\multicolumn{7}{c}{Galaxy Number in Each Field} \\
\cline{3-9}
 & & 
\multicolumn{3}{c}{$\log F_{\ha}$\footnotemark[$*$]$ \geq -16.0$} & \hspace{1pt} & 
\multicolumn{3}{c}{$\log F_{[\rm O \emissiontype{II}]}$\footnotemark[$*$]$ \geq
-16.5$}\\ 
 & & $z$=0.5--1.0 & $z$=1.0--1.4 & $z$=1.4--1.7 
& & $z$=0.5--1.0 & $z$=1.0--1.4 & $z$=1.4--1.7 \\ 
\hline
\hline
\endhead
SXDF 1 & 0.179 & 1130 & 1361 & 386 & & 1523 & 1238 & 331 \\ 
SXDF 2 & 0.199 & 1235 & 1311 & 371 & & 1768 & 1241 & 322 \\ 
SXDF 3 & 0.176 & 1119 & 1044 & 419 & & 1610 & 1005 & 358 \\ 
SXDF 4 & 0.178 & 1134 & 907 & 394 & & 1691 & 850 & 328 \\ 
\hline 

$\sigma_{\rm gal, obs}$ \footnotemark[$\dagger$] & & 
0.012 & 0.168 & 0.099 && 0.047 
& 0.153 & 0.093 \\
$ \Delta \sigma_{\rm gal, obs}$ \footnotemark[$\ddagger$] & & 
$\pm$0.009 & $\pm$0.137 & $\pm$0.081 && $\pm$0.038 
& $\pm$0.124 & $\pm$0.075 \\
$b \sigma_{\rm CDM}$ \footnotemark[$\S$] & & 
0.125 & 0.105 & 0.104 & & 0.120 & 0.103 & 0.109 \\
\hline
\multicolumn{9}{@{}l@{}}{\hbox to 0pt{\parbox{170mm}{
\footnotesize 
\par\noindent 
\footnotemark[$*$] In units of line flux [$\flux$]. 
\par\noindent 
\footnotemark[$\dagger$] The standard deviation of galaxy density
fluctuation, $\delta \rho_{\rm gal} / \bar{\rho}_{\rm gal}$ estimated
from the observed galaxy numbers in sub-fields.
\par\noindent 
\footnotemark[$\ddagger$] The observational error of $\sigma_{\rm gal, obs}$.
\par\noindent 
\footnotemark[$\S$] The expected density fluctuations predicted from the
CDM theory and the bias estimates, which should be compared with
$\sigma_{\rm gal, obs}$.@
}\hss}}
\end{longtable}

We then estimate the standard deviation $\sigma_{\rm gal, obs}$ of the
galaxy density fluctuation $\delta_{\rm gal} = \delta \rho_{\rm gal} /
\bar{\rho}_{\rm gal}$ on the volume scale of the sub-fields as
follows.  When the surface galaxy number density $n_i$ in the $i$-th
sub-field obeys a distribution with the mean value $\bar{n}$, the
unbiased estimate for the standard deviation $\sigma_n$ of the
distribution from the observed number densities is given by:
\begin{eqnarray} 
\sigma^2_n = 
\frac{1}{N-1} \sum_{i=1}^N ( n_i - \bar{n}_{\rm obs} )^2 \ ,
\end{eqnarray} 
where $\bar{n}_{\rm obs} = \sum n_i / N$ is the observational estimate
of the mean value. Then, $\sigma_{\rm gal, obs}$ is simply estimated
by $\sigma_n / \bar{n}_{\rm obs}$. The observational error in the
$\sigma_{\rm gal, obs}$ estimate can be calculated by approximating
that $\sum (n_i - \bar{n}_{\rm obs})^2 / \sigma_n^2$ obeys to $\chi^2$
distribution with $N-1$ degrees of freedom.  
The value of $\sigma_{\rm
  gal, obs}$ and its error are thus estimated for galaxies brighter
than $10^{-16} \flux$ (H$\alpha$) and $10^{-16.5}
\flux$ ($[\rm O \emissiontype{II}]$), which are given in Table
\ref{field-to-field} in the three redshift ranges.

To compare with these results, we calculate the theoretically expected
standard deviations, $b\sigma_{\rm CDM}$, where $\sigma_{\rm CDM}$ is
the standard deviation of dark matter fluctuation and $b$ is the
linear bias of line emitting galaxies.  The value of $\sigma_{\rm
  CDM}$ is calculated assuming a spherical top-hat window function
with radius $R$, and $R$ is assumed to be $R = \left( 3V/4\pi
\right)^{\frac{1}{3}}$, where $V$ is the averaged volume of the four
sub-fields of SXDF. The results are given in Table
\ref{field-to-field}, to be compared with $\sigma_{\rm gal, obs}$. The
values of $b\sigma_{\rm CDM}$ are in agreement with $\sigma_{\rm gal,
  obs}$ in $1.0 < z <1.4$ and $1.4 < z < 1.7$, but in disagreement in
$0.5 < z < 1.0$.  The origin of the discrepancy only in $0.5 < z <
1.0$ is not clear, but the estimate of $\sigma_{\rm gal, obs}$ is based
only on four subfields.  We find that the chance probability of
getting this level of deviation is about 4\% and 20\% for $\ha$ and
[O\emissiontype{II}] line cases, respectively, and hence the
possibility of a statistical fluke cannot be excluded.


\section{Example of Color Selection}\label{section: color selection}

Actual selection procedures in a particular BAO survey will depend on
various conditions that are unique to the survey, such as available
band filters of input imaging surveys. Therefore it is difficult to
derive generally useful results, but here we test some simple
two-color selection methods to select emission line galaxies brighter
than $F_{\ha} \geq 10^{-16} \flux$: one is using only optical
photometries ($Bi'z'$), and the other is including the NIR $K$ band
($Bz'K$). We do not show in detail the results for [O\emissiontype{II}], 
but we have confirmed that the results are similar when one
considers the corresponding [O\emissiontype{II}] line flux of
10$^{-16.5}$ $\flux$.

\begin{figure}
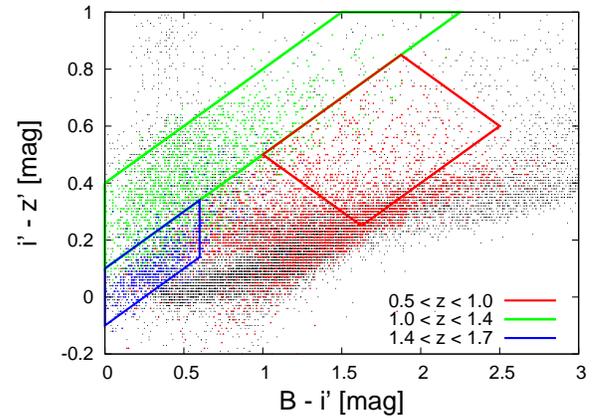

\begin{center}
\FigureFile(80mm,80mm){fig18_Biz.eps}
\end{center}
\caption{ The $Bi'z'$ plot for SXDF galaxies with $B < 24$.  Galaxies
  with $\ha$ flux brighter than $10^{-16}$ $\flux$ in the
  three redshift ranges are indicated by red, green, and blue dots,
  while the other galaxies in black dots. The color selection criteria
  used for the three redshift ranges are indicated by the
  corresponding colors. 
}
\label{Biz} 
\end{figure}

\begin{figure}
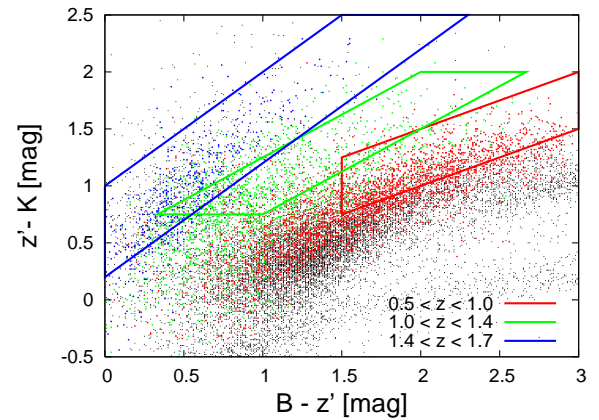

\begin{center}
\FigureFile(80mm,80mm){fig19_BzK.eps}
\end{center}
\caption{
The same as Figure \ref{Biz}, but for $Bz'K$ diagram.
} 
\label{BzK} 
\end{figure}

Before we examine the two-color diagrams, we should select relatively
bright galaxies as potential targets for BAO surveys, using
photometric information.  From the magnitude distribution of
emission-line galaxies (Figure \ref{Ha_magfunc}), we set the magnitude
threshold as $B < 24$.  We then show the color-color diagrams of all
the SXDF galaxies with $B < 24$ for the two cases, in Figure \ref{Biz}
and Figure \ref{BzK}. Galaxies that are brighter than $F_{\ha} =
10^{-16} \flux$ are shown by colored dots, for the three different
redshift intervals.  From these results, we define the color selection
criteria to efficiently select the emission-line galaxies, as shown by
the box in the same colors for the three redshift intervals.  These
criteria were chosen not only by the number of target emission line
galaxies, but also by the expected success rate examining the
contamination of non-emission line galaxies (black dots).  These
criteria can also be applied for [O\emissiontype{II}] emitters
brighter than 10$^{-16.5}$ $\flux$ because the distributions on the
two-color planes are similar to those in the H$\alpha$ cases.

\begin{figure*}
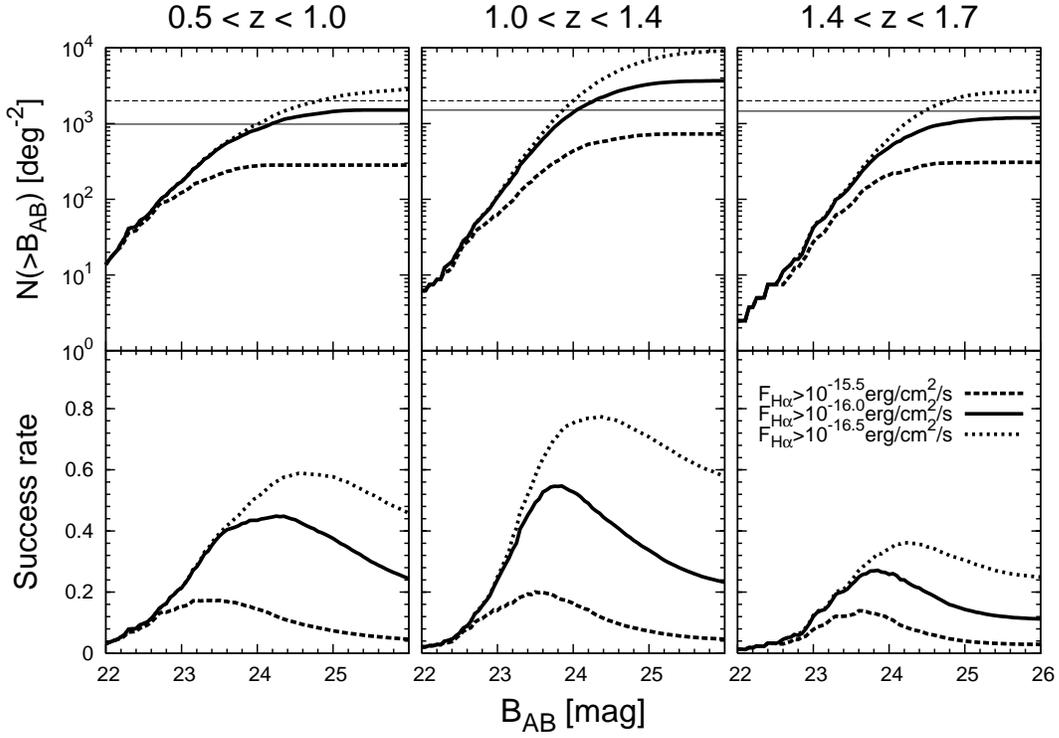

\begin{center}
\FigureFile(140mm,140mm){fig20_Biz_color_select.eps}
\end{center}
\caption{ The counts and probability of successful selection in the
  three redshift intervals by the $Bi'z'$ selection method, for
  galaxies brighter than a given threshold $B$ magnitude
  (abscissa). The galaxy sample is SXDF.  The horizontal lines (solid
  and dashed) in the top panels indicate the surface number density
  corresponding to $nP = 1$ at each redshift interval and the FMOS
  fiber density, respectively. The dashed, solid, and dotted lines are
  for the three threshold $\ha$ fluxes of $10^{-15.5}$, $10^{-16}$,
  and $10^{-16.5} \flux$, respectively.  }
\label{Biz_success}
\end{figure*}

\begin{figure*}
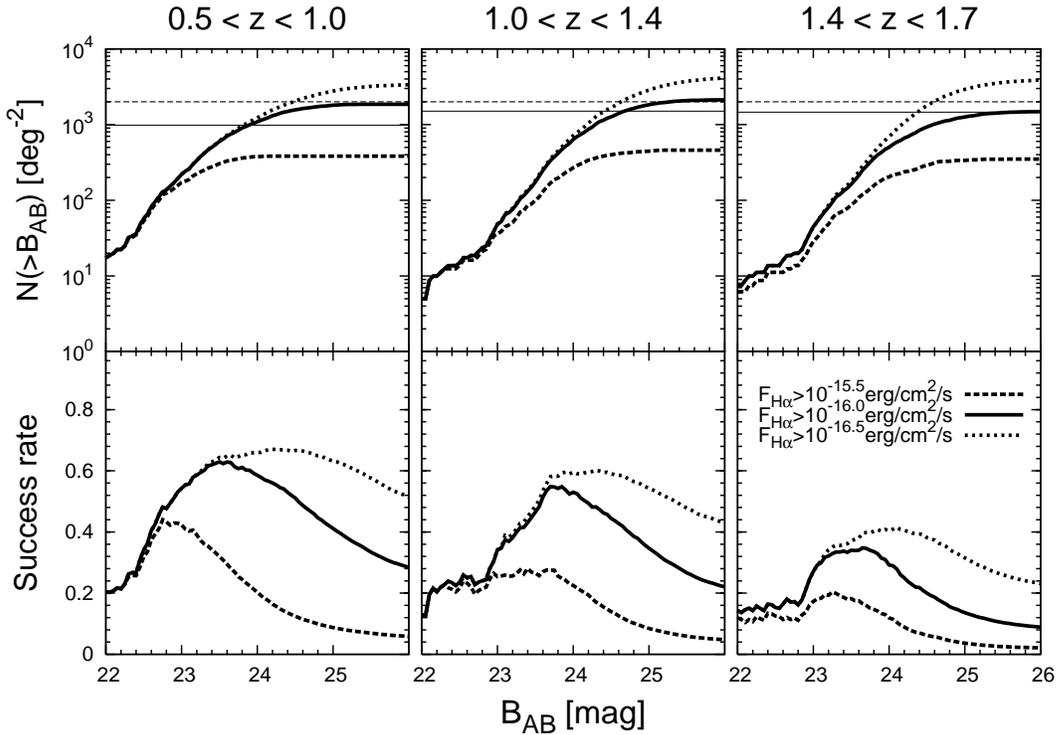

\begin{center}
\FigureFile(140mm,140mm){fig21_BzK_color_select.eps}
\end{center}
\caption{The same as Fig. \ref{Biz_success}, but for the $Bz'K$ selection.
}
\label{BzK_success}
\end{figure*}

\begin{figure*}
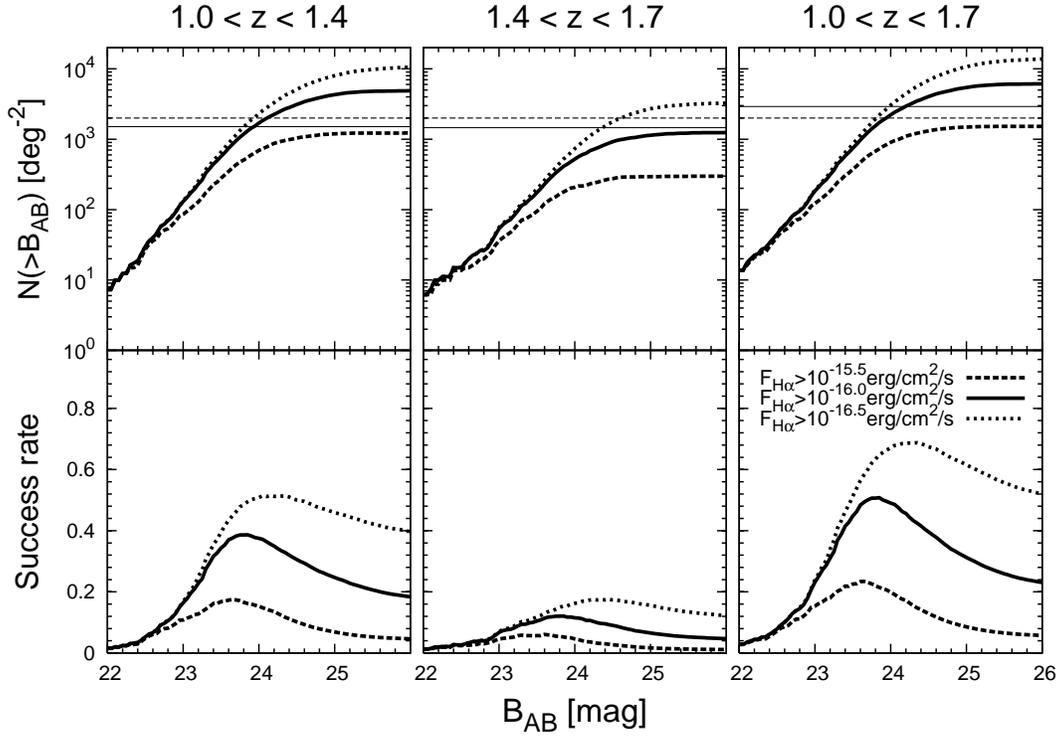

\begin{center}
\FigureFile(140mm,140mm){fig22_Biz_color_select_z1017.eps}
\end{center}
\caption{The same as Fig. \ref{Biz_success}, but for the $Bi'z'$ selection on 
  the $1.0 < z < 1.7$ joint condition. 
  The left, middle, and right panels show the counts and probability 
  of the successful selection for galaxies 
  in the redshift intervals of $1.0<z<1.4$, $1.4<z<1.7$, and
  in the joint redshift interval of $1.0<z<1.7$.
}
\label{Biz_success_z1017}
\end{figure*}

\begin{figure*}
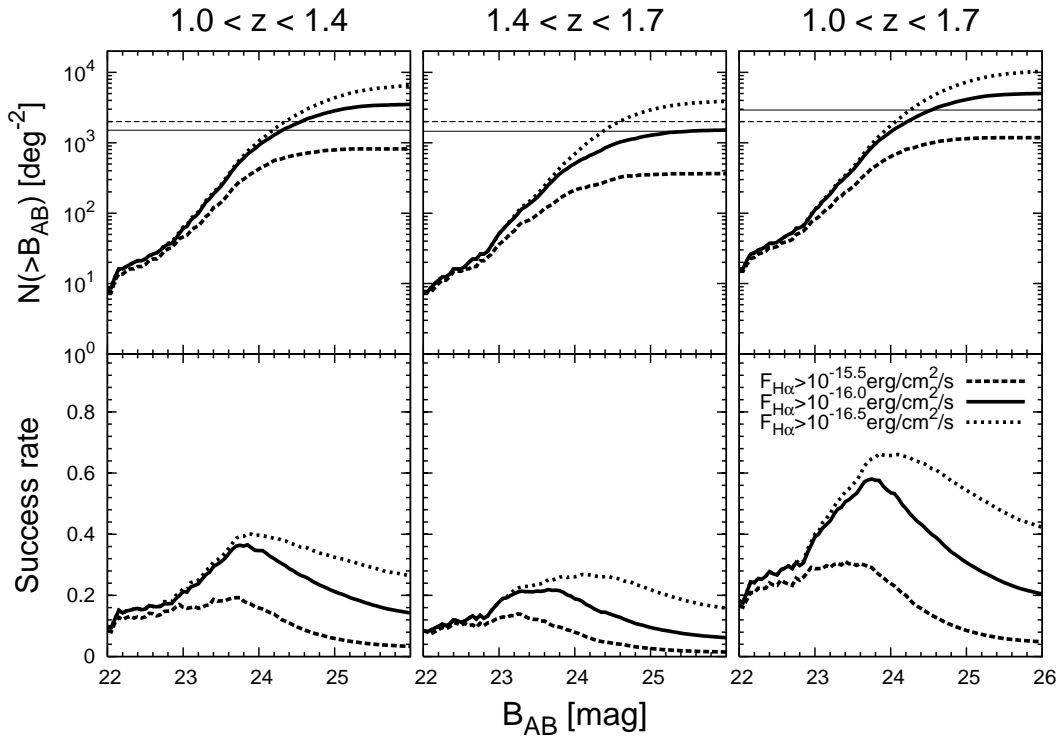

\begin{center}
\FigureFile(140mm,140mm){fig23_BzK_color_select_z1017.eps}
\end{center}
\caption{The same as Fig. \ref{Biz_success_z1017}, but for the $Bz'K$ selection on 
  the $1.0 < z < 1.7$ joint condition. 
}
\label{BzK_success_z1017}
\end{figure*}

Figures \ref{Biz_success} and \ref{BzK_success} show the number of
emission-line galaxies brighter than three threshold $\ha$ line fluxes
of $10^{-15.5}$, $10^{-16}$, and $10^{-16.5} \flux$, as a function of
the threshold $B$ magnitude when we select galaxies by these color
criteria. It can be seen that for the two low redshift bins sufficient
galaxies ($nP \gtrsim 1$) are obtained at a depth of $B=24$ with a
$10^{-16} \flux$ threshold.  For the highest redshift bin we need to
go deeper ($B=25$) to attain sufficient galaxies.

The success rate of the selection, i.e., the fraction of galaxies
having larger emission line flux than the threshold and in the correct
redshift range is also shown in these figure, compared with the total
number of all galaxies satisfying the $B$ magnitude threshold and
color selection criteria. These results indicate that typically a
success rate of about 40\% in the two low redshift bins can be
achieved by this simple color selection for the threshold line flux of
$F_{\ha} = 10^{-16} \flux$.  For the case for [O\emissiontype{II}], we
also find that typically a success rate of about 40\% in the two low
redshift bins can be achieved for the threshold line flux of 
$F_{\rm [O\emissiontype{II}]} = 10^{-16.5} \flux$.

Finally we consider the two-color selection for high redshift galaxies at $z =$ 1--1.7
using $Bi'z'$ and $Bz'K$ photometries. This may be effective if we do not discriminate galaxies 
beyond or below $z=1.4$ and are simply interested in selecting galaxies at $z>1$. 
We define the selection criteria for $Bi'z'$ and $Bz'K$ as shown by the two high redshift boxes 
in Figure \ref{Biz} and \ref{BzK}, respectively. 
From Figure \ref{Biz_success_z1017} and \ref{BzK_success_z1017}, 
we can select a sufficient number of galaxies at $1.0 < z < 1.7$ by a survey deeper 
than $B=24$ with a success rate of 50--60\%. 
If the survey depth is deeper than $B=25$, we can select a sufficient number of galaxies 
both in the two redshift intervals of $1.0 < z < 1.4$ and $1.4 < z < 1.7$.

We find that both $Bi'z'$ and $Bz'K$ selection deliver comparable
performance for selection in terms of magnitude depth required and
success rate. The choice would then come down to the ease of obtaining
the observational datasets.

Possible improvements would be to try three-color selection boxes, or
multi-parameter methods in an N-dimensional photometric space. These
improvements are beyond the scope of this paper but we will explore
such possibilities in future work.


\section{Summary}\label{section: summary}

We made an estimate of H$\alpha$ and [O\emissiontype{II}] emission
line luminosities of SDF and SXDF galaxies in a total field area of
0.846 deg$^2$ by a photometric redshift based approach, for the purpose of
studying photometric selection procedures of high redshift emission
line galaxies useful for future spectroscopic BAO surveys. Our line
luminosity estimates were tested and calibrated by using the SDSS
galaxies with spectroscopic line luminosity measurements at low
redshifts. The line luminosity functions of H$\alpha$ and
[O\emissiontype{II}] were thus derived in three reshift intervals in
$z$ = 0.5--1.7, and they are in reasonable agreement with the previous
studies based on spectroscopic or narrow-band filter observations.

We examined the number density and properties of emission line
galaxies that can be used for future BAO surveys, taking FMOS and
WFMOS instruments for the Subaru Telescope as representative examples.
We have confirmed that there are sufficient number of emission line galaxies
to perform BAO surveys in the redshift ranges $0.5<z<1$ and $1<z<1.4$ 
which are bright enough to be detected by 8--10 m
class telescopes with exposures shorter than one hour.
We showed cumulative distributions of magnitudes in popular
band filters for emission line galaxies with line flux brighter than
given thresholds in a range of redshift intervals, and these can be used to
estimate the required depth of an imaging survey from which
spectroscopic targets are selected. 

We also estimated the stellar mass distributions,  extinction distributions, 
FWHM size distributions, and linear biases of line emitting galaxies. The linear
biases were inferred from the stellar mass and redshift estimates,
combined with the standard theory of structure formation. The inferred
biases are consistent with the field-to-field variation of galaxy
numbers in four subfields in SXDF, and also with the estimates based
on clustering analysis for similar galaxies in previous studies.

Finally, we showed some examples of color selection of target galaxies
for spectroscopic BAO surveys, based on simple color-color selections
($Bi'z'$ and $Bz'K$). It is found that target galaxies for BAO surveys
can be selected with a reasonable accuracy and success rate.  
We find that with this selection for $z<1.4$ we need  to reach
a typical depth of $B=24$ (with other filters being required for color-selection) 
and for $1.4<z<1.7$ we need to reach  $B=25$. We find the
$Bi'z'$ and $BzK'$ selection methods give comparable efficiency. In
general we find that it is not possible to efficiently select galaxies in the high-redshift
 $1.4<z<1.7$ using only two-colour selection. 

These results give guidance for the planning of future BAO surveys and
demonstrate that 8-m telescopes with spectroscopic fields of view of order 
one degree (instruments such as FMOS and WFMOS) would be effective machines 
for carrying out such large cosmological surveys.

The data of photometric magnitudes, redshifts, and line luminosities used 
in this work are available on request to the authors.

\vspace{\baselineskip}

We would like to thank K. Harikae for useful discussions and
E. Wisnioski for providing us with the observational results of
WiggleZ survey.  This work was supported by the Global COE Program
"The Next Generation of Physics, Spun from Universality and Emergence"
and Grant-in-Aids for Scientific Research (Nos. 19740099 and 20040005)
from the Ministry of Education, Culture, Sports, Science and
Technology (MEXT) of Japan.  This work was also supported by the Japan
Society for Promotion of Science (JSPS) Core-to-Core Program
`International Research Network for Dark Energy'.  Karl Glazebrook
acknowledges support for this work from Australian Research Council
(ARC) Discovery Projects DP0774469 and DP0772084.


\end{document}